\documentclass[aps,prr,amsmath,amssymb,floatfix,twocolumn,superscriptaddress,10pt]{revtex4-1}
\usepackage{graphicx}
\usepackage{subfigure}
\usepackage{hyperref}
\usepackage{braket}
\usepackage{bm}
\usepackage{comment}
\usepackage{xcolor}
\usepackage{soul}
\setstcolor{red}

\DeclareMathAlphabet\mathbfcal{OMS}{cmsy}{b}{n}

\def\YBCO248{YBa$_2$Cu$_4$O$_8$}

\def\br{{\bf r}}

\def\bq{{\bf q}}

\def\bq{{\bf q}}
\def\bQ{{\bf Q}}

\def\prl{{Phys. Rev. Lett. }}
\def\prb{{Phys. Rev. B }}
\def\prx{{Phys. Rev. X }}

\newcommand\redsout{\bgroup\markoverwith{\textcolor{red}{\rule[0.5ex]{2pt}{0.4pt}}}\ULon}

\makeatletter
\newsavebox{\@brx}
\newcommand{\llangle}[1][]{\savebox{\@brx}{\(\m@th{#1\langle}\)}%
  \mathopen{\copy\@brx\kern-0.5\wd\@brx\usebox{\@brx}}}
\newcommand{\rrangle}[1][]{\savebox{\@brx}{\(\m@th{#1\rangle}\)}%
  \mathclose{\copy\@brx\kern-0.5\wd\@brx\usebox{\@brx}}}
\makeatother

\begin{document}

\title{Quantum Monte Carlo study of a bilayer $U(2)\times U(2)$ symmetric Hubbard model}

\author{Yosef~Caplan}
\affiliation{Racah Institute of Physics, The Hebrew University,
  Jerusalem 91904, Israel}
\author{Dror~Orgad}
\affiliation{Racah Institute of Physics, The Hebrew University,
  Jerusalem 91904, Israel}

\date{\today}

\begin{abstract}

We carry out a sign-problem-free quantum Monte Carlo calculation of a bilayer model
with a repulsive intra-layer Hubbard interaction and a ferromagnetic inter-layer
interaction. The latter breaks the global $SU(2)$ spin rotational symmetry but
preserves a $U(2)\times U(2)$ invariance under mixing of same-spin electrons
between layers. We show that despite the difference in symmetry, the bilayer model exhibits
the same qualitative features found in the single-layer Hubbard model. These include
stripe phases, whose nature is sensitive to the presence of next-nearest-neighbor hopping,
a maximum in the Knight shift that moves to lower temperatures with increasing hole doping,
and lack of evidence for intra-layer $d$-wave superconductivity.
Instead, we find a superconducting phase, coexisting with stripes, whose critical temperature traces a dome
as a function of doping and is due to inter-layer spin-polarized pairing that is
induced by the ferromagnetic interaction.

\end{abstract}

\maketitle

\section{Introduction}

Establishing the properties of strongly interacting models, especially in dimensions larger than one,
is a difficult problem. A canonical example is the two-dimensional fermionic Hubbard model \cite{Hubbard-review},
whose apparent simplicity and widely believed relevance to the high-temperature superconductors have motivated
an enormous amount of work over the past six decades. Still, apart from the half-filled system with $n=1$ electrons
per site \cite{half-filled}, the weakly interacting limit $U/t\rightarrow 0$, where $U$ and $t$ are respectively
the on-site repulsion and inter-site hopping \cite{Hubbard-RG}, and the Nagaoka limit $U/t\rightarrow\infty$ with
a single doped hole \cite{Nagaoka}, not much is known with theoretical confidence. Particularly challenging is the
intermediate range $U\sim t$, where obtaining a faithful map of the model's behavior depends on numerical calculations.

Currently, the leading technique to study this regime, both in terms of its reliability and ability to handle
relatively large systems, is the density matrix renormalization group (DMRG). To date, DMRG has been used to
study Hubbard  cylinders with up to six legs
\cite{6leg-Hubbard-2003,Noack-Hubbard-stripes,6leg-Hubbard-Hager,Stripes-Hubbard-comapre,4leg-Hubbard-Science,
4leg-Hubbard-cylinder,4leg-PRR,Hubbard-absence,coexistence,6legHubbard},
where typically $U/t=8-12$ and $n=0.875$. The consequences of including next-nearest-neighbor hopping $t'$
were also addressed \cite{4leg-Hubbard-Science,4leg-Hubbard-cylinder,4leg-PRR,coexistence,6legHubbard}.
More numerous are DMRG studies of the $t$-$J$ model - the large $U/t$ descendent of the Hubbard model,
on cylinders with up to eight legs
\cite{4leg-PRR,ttpJ-White-1999,ttpJ-White-2009,ttpJ-White-2012,intertwined-4leg-tJ,4leg-tJ-SC,6leg-tJ-spin-liquid,
6leg-tJ-Gong,8leg-tJ-Jiang,8leg-tJ-Lu}.
In most cases the calculations were carried out for $J/t=1/3$, which would correspond to $U/t=12$ if the mapping
to the Hubbard model holds down to this range of interaction strengths, and for hole densities of up to 1/8.
Several studies included $t'$-hopping, which at times was also accompanied by a $J'$ term
\cite{intertwined-4leg-tJ,6leg-tJ-spin-liquid,6leg-tJ-Gong}.

The findings of these studies may be roughly summarised as follows: (i) The vicinity of $t'=0$ is characterized
by charge-density wave (CDW) modulations in the form of filled stripes with one hole per unit length domain wall
\cite{Noack-Hubbard-stripes,Stripes-Hubbard-comapre,4leg-PRR,Hubbard-absence,6leg-tJ-Gong}
(nearly half-filled stripes or with 2/3 filling were also observed \cite{6leg-Hubbard-Hager,6legHubbard,4leg-tJ-SC},
and the various types are almost degenerate \cite{Stripes-Hubbard-comapre}). They are accompanied by short-ranged
spin-density wave (SDW) modulations with twice the period \cite{4leg-tJ-SC,4leg-PRR}
and by exponentially decaying $d$-wave superconducting ($d$-SC) correlations \cite{4leg-PRR,Hubbard-absence}
(see, however, Ref. \cite{4leg-tJ-SC}).
(ii) The presence of $t'<0$ causes the stripes to become half-filled \cite{4leg-Hubbard-cylinder,4leg-PRR,8leg-tJ-Jiang}
or exhibit an intermediate filling between 0.5 and 1 \cite{coexistence}. On four-leg cylinders both the CDW
and the $d$-SC correlations decay as power-laws, but the former dominate. There are conflicting results
on wider systems. While only short-range $d$-SC correlations have been found on a six-leg
cylinder \cite{6legHubbard}, non-zero $d$-SC order was also reported \cite{coexistence}.
Regardless, the SDW correlations are still modulated with twice the CDW period and decay exponentially.
(iii) For $t'>0$ and larger than a small threshold the system enters a phase with no stripes and robust
power-law $d$-SC correlations \cite{6leg-tJ-Gong,8leg-tJ-Jiang,8leg-tJ-Lu}. Increasing $t'$ further makes
partially-filled stripes reappear. The power-law superconducting correlations decay somewhat faster than
the CDW correlations in Hubbard cylinders, while the situation is reversed for the $t$-$J$ model
\cite{4leg-PRR,6legHubbard,6leg-tJ-Gong}. In both cases the spin correlations decay exponentially.

Notwithstanding its advantages, DMRG is largely limited to ladder geometries as it involves a computational cost
that grows exponentially with the ladder width. Furthermore, it provides information about the ground state, and
using it to extract dynamical or finite-temperature information is still in an early stage. Hence, it is desirable
to augment DMRG by another method that allows to probe more two-dimensional geometries away from the strict
zero-temperature limit. To this end, the determinant quantum Monte Carlo (DQMC) technique appears as the method
of choice. Like DMRG it is also unbiased and, in principle, numerically exact. However, away from half filling
it is plagued by the sign problem that incurs a prohibitive computational cost as one attempts to explore
temperatures much smaller than the bandwidth. Nevertheless, several "brute force" unconstrained DQMC studies
\cite{QMC-3band-Hubbard-stripes,QMC-Hubbard-stripes,QMC-Hubbard-JPSJ,QMC-Hubbard-Yang}
were able to probe the model down to temperatures of about $T\approx 0.2 t$. Their findings show that even
at these relatively high temperatures the Hubbard model exhibits ubiquitous and robust stripy correlations,
in agreement with the DMRG results. At the same time, no signs of superconductivity were detected.

Here, we pursue a complementary approach where we use DQMC to study a model that is free of the sign problem,
as a computational proxy to  the two-dimensional Hubbard model. Specifically, we revisit a bilayer model that
was originally introduced by Assaad et al. \cite{Assad-bilayer}, describing two Hubbard layers that are further
coupled by a ferromagnetic interaction between neighboring sites belonging to different layers. While Ref.
\cite{Assad-bilayer} considered only ground-state stripes correlations for few doping levels and $t'=0$,
we have calculated various charge, spin and superconducting finite-temperature correlation functions over
a wide doping range and included the effects of next-nearest-neighbor hopping. Our goal is to contrast the
behavior of the bilayer model with the available data on the single-layer Hubbard model in order to establish
the level at which the former may be used to glean insights about the latter. This is not a priori clear since
the ferromagnetic inter-layer coupling breaks the global $SU(2)$ spin rotation symmetry of the Hubbard model.
Concomitantly, it leaves intact a $U(2)\times U(2)$ symmetry, where each $U(2)$ transformation mixes same-spin
electrons between the two layers.

Our findings show strong similarities between the electronic signatures of the two models.
In particular, the four-leg bilayer sustains spin and charge stripe phases whose dependence on $t'$
and electronic density follows closely that of stripes in the corresponding Hubbard system, as outlined
above. The overall trends persist also in the square systems that we have investigated.
When $t'=0$ we find filled charge stripes and spin stripes whose correlation length is larger than the
accessible system sizes up to a hole-doping level of about 0.25, from where it steadily decreases.
For $t'=-0.25t$, stripes exist over the same doping range but the charge stripes host only 4/5-2/3 holes per
unit length of the domain wall, in close resemblance to a recent DMRG study of a six-leg
Hubbard cylinder \cite{6legHubbard}. For $t'=0.25t$, the square systems exhibit fractionally filled stripes
and only above a minimal hole concentration that resides near 1/8. However, we can not rule out their
existence at lower doping levels in the thermodynamic limit.

We have looked for signatures of intra-layer $d$-wave superconductivity by calculating the corresponding
susceptibility and vertex function. Our findings for $t'=-0.25t$ and temperatures above $T=0.2t$ conform with
a DQMC study of the Hubbard model under similar conditions \cite{QMC-Hubbard-JPSJ}, which did not provide
any evidence for a $d$-SC instability. Extending the search down to $T=0.05t$ did not change this conclusion,
nor did changing the sign of $t'$. Despite the fact that the largest values of both the susceptibility
and the vertex function were obtained for $t'=0.25t$ and below 0.2 hole doping, neither show signs of the
finite size scaling expected from the onset of $d$-SC order. In contrast, we did find a sharp rise in the
superconducting stiffness at low temperatures to values above the threshold for a Berezinskii-Kosterlitz-Thouless
(BKT) transition. We provide evidence that the resulting superconducting phase coexists with stripes
and is due to inter-layer spin-polarized pairing induced by the ferromagnetic interaction. Finally, the uniform
spin susceptibility (Knight shift) peaks at a temperature $T^*$ that decreases with increasing doping,
as previously found for the Hubbard model \cite{QMC-Hubbard-JPSJ}. Our ability to probe the bilayer model
down to much lower temperatures allows us to detect the leveling off of $T^*$ above 0.3 hole doping.

\section{Model and Methods}

To ensure that a fermionic Hamiltonian is free of the sign problem it is sufficient that its kinetic
part and its Hubbard-Stratonovich-decoupled interaction commute with some antiunitary operator \cite{Zhang-sign}.
A special case is when the fermionic determinant factorizes into two identical real copies, thus guaranteeing
its positivity. Pursuing this route, Assaad et al. \cite{Assad-bilayer} considered the following bilayer
Hamiltonian on a square lattice, which we also study
\begin{eqnarray}
\label{eq:bilayerH}
\nonumber
H&=&-\sum_{l=1,2}\sum_{\sigma=\uparrow,\downarrow}\left(\sum_{i,j}t_{ij}c_{li\sigma}^\dagger c_{lj\sigma}
+\mu\sum_i c_{li\sigma}^\dagger c_{li\sigma} \right) \\
&&-\frac{U}{4}\sum_i \left(n_{1i\uparrow}-n_{1i\downarrow}+n_{2i\uparrow}-n_{2i\downarrow} \right)^2.
\end{eqnarray}
Here, $l$ is the layer index, $\mu$ is the chemical potential and $n_{li\sigma}=c_{li\sigma}^\dagger c_{li\sigma}$.
The hopping amplitudes take the value $t$ between neighboring sites within a layer, and $t'$ between next-nearest
neighbors on the same layer. Throughout the paper we use a unit lattice constant and set $t=1$, which serve as
the basic length and energy scales. The interaction is also expressible as $-(U/4)\sum_{li\sigma}n_{li\sigma}
+(U/2)\sum_{li}n_{li\uparrow}n_{li\downarrow} -2U\sum_i S^z_{1i}S^z_{2i}$, where $S^z_{li}=(n_{li\uparrow}-n_{li\downarrow})/2$.
Hence, up to a shift of the chemical potential it amounts to local Hubbard repulsion (assuming $U>0$) on each layer
and a ferromagnetic coupling between neighboring sites on different layers. Note that the latter acts to penalize
double occupancy on either layers and thus adds to the effective Hubbard repulsion.
More importantly, while the interaction is invariant under $U(2)\times U(2)$ transformations acting separately
on the two subspaces of same-spin electrons, it breaks the global spin rotation symmetry and introduces effective
attraction between the layers.

The particle-hole transformation, $c_{li\sigma}\rightarrow (-1)^i c_{li\sigma}^\dagger$,
where the factor $(-1)^i$  equals -1 on one sublattice and 1 on the other, leaves the
Hamiltonian invariant with the exception that $t'\rightarrow -t'$. It also changes the average
site occupation according to $\langle n\rangle\rightarrow 2-\langle n\rangle$. Hence, we concentrate
on the hole-doped regime $\delta=1-\langle n \rangle >0$, and rely on the relation between the expectation
values of observables $\langle{\cal O}\rangle(t',\delta)=\langle{\cal O}\rangle(-t',-\delta)$ to deduce
the behavior in the electron-doped regime from the hole-doped counterparts.

\begin{figure*}
\centering
\subfigure{\includegraphics[width=0.325\textwidth,height=0.23\textwidth]{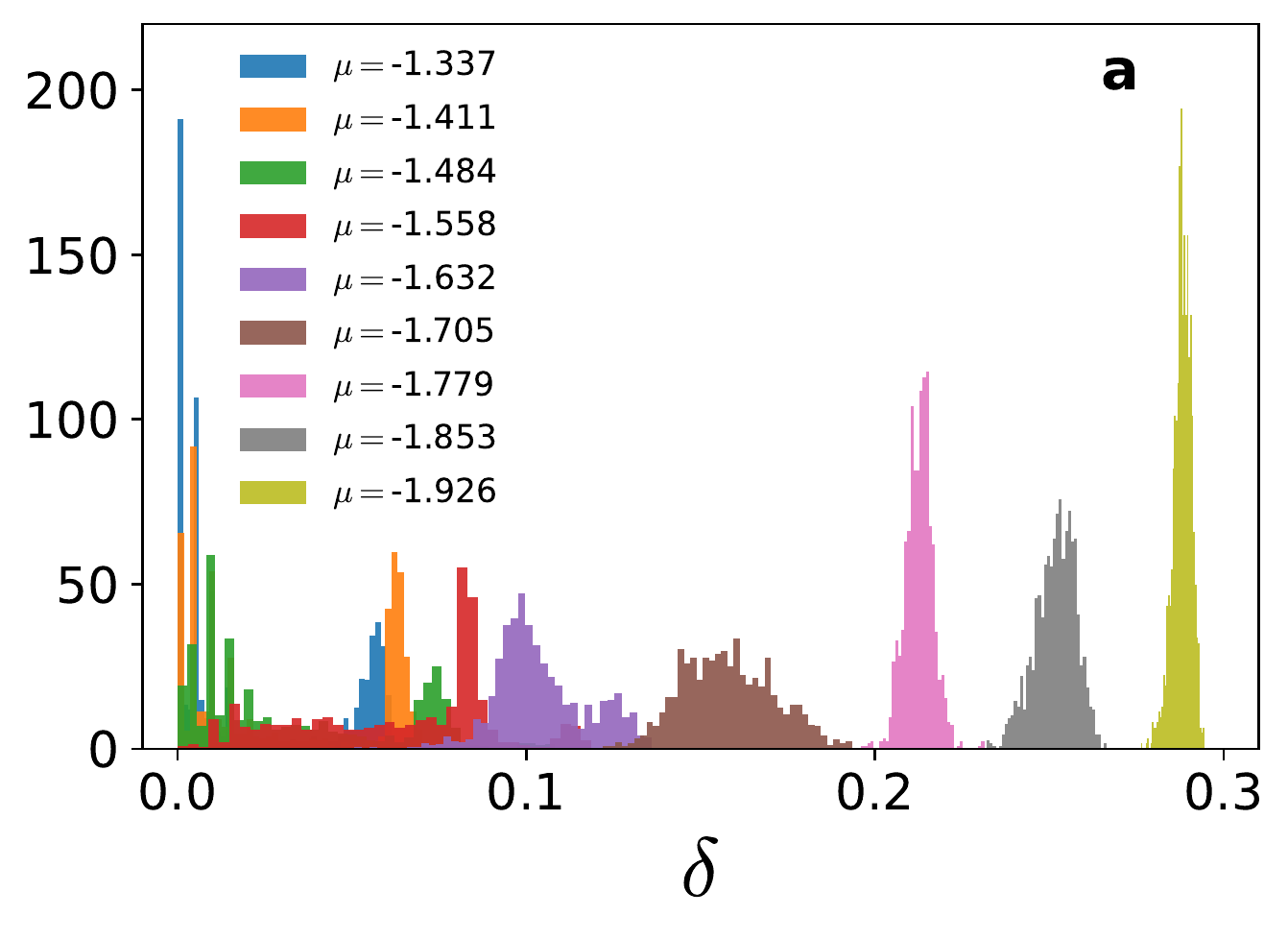}}
\subfigure{\includegraphics[width=0.325\textwidth,height=0.23\textwidth]{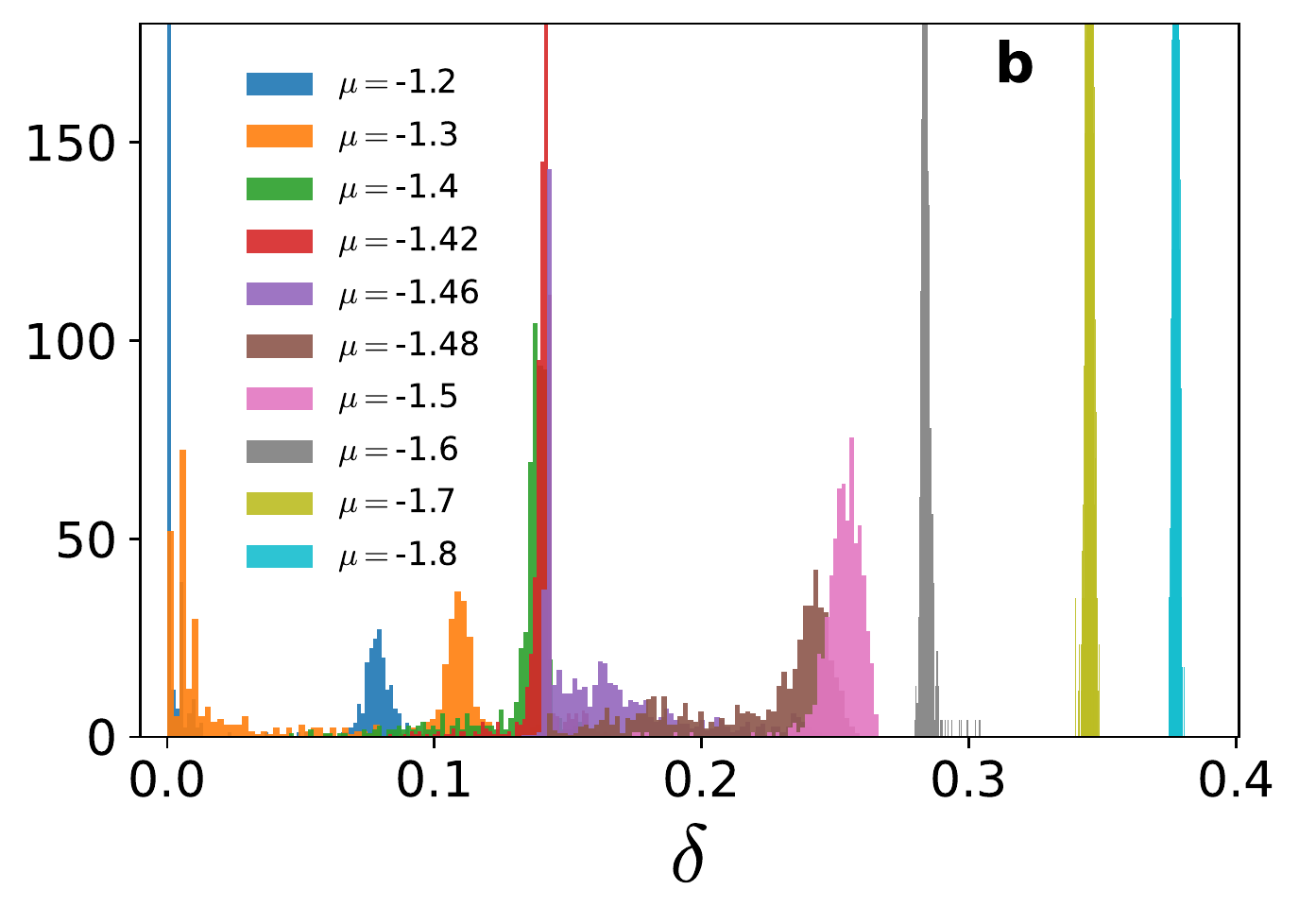}}
\subfigure{\includegraphics[width=0.325\textwidth,height=0.23\textwidth]{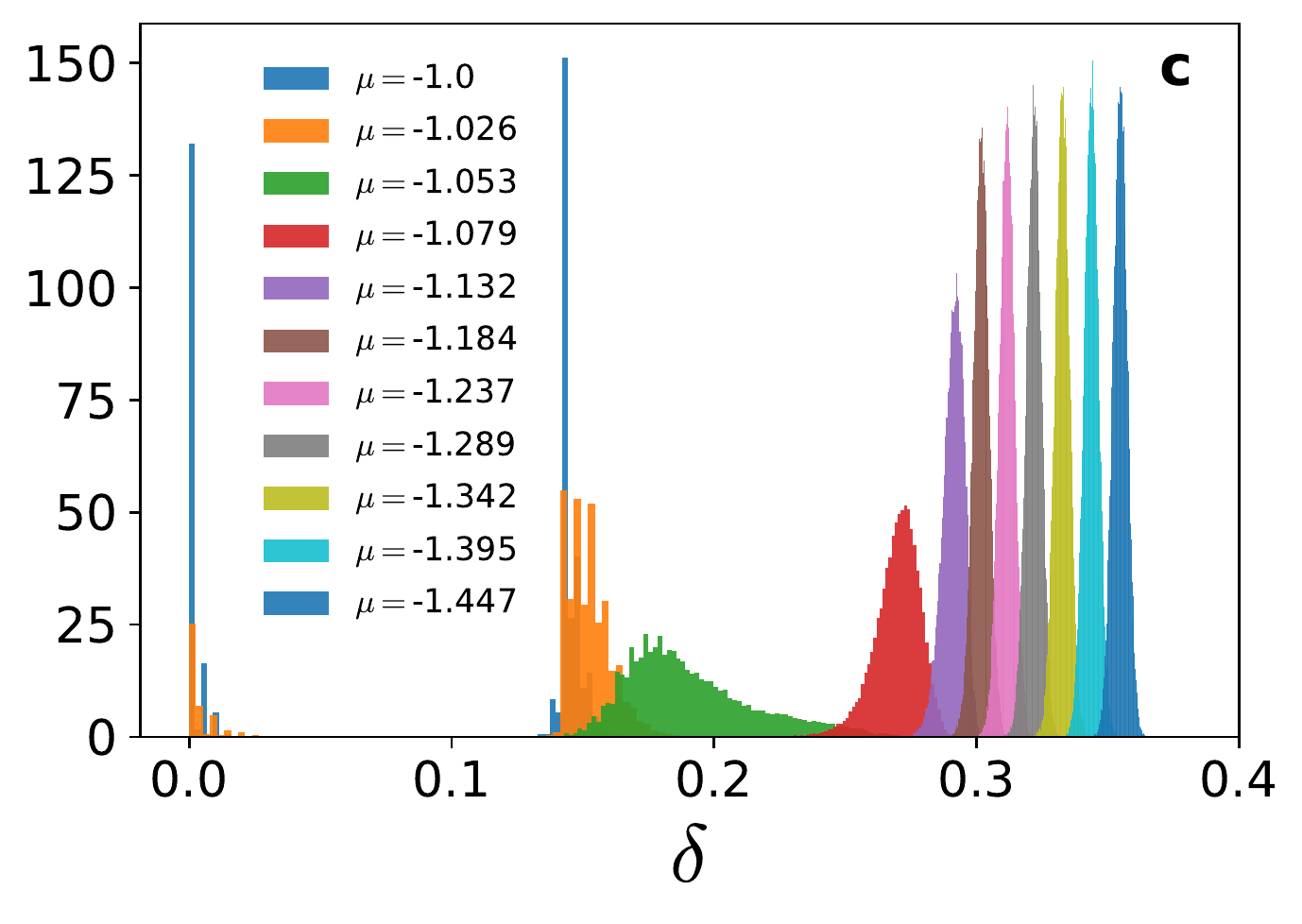}}
\caption{Histogram of the hole concentration $\delta$ obtained from Monte Carlo sampling of a $14\times 14$
square bilayer with (a) $t'=-0.25$, (b) $t'=0$, and (c) $t'=0.25$, for various values of the chemical
potential $\mu$ at a temperature $T=0.05$.}
 \label{fig:PS}
\end{figure*}

In order to use the DQMC method we decouple the interaction term in the Hamiltonian via
a discrete Hubbard-Stratonovich transformation that involves two
Ising-like fields taking values $\pm\psi_{1,2}$. They are accompanied by coefficients
$a_m$ \cite{Assaad-HS}, which are chosen to ensure the validity of
 \begin{equation}
e^{\Delta \tau U\tilde{n}^2}=\frac{1}{4}\sum_{m=1,2}\sum_{\lambda=\pm 1}a_m e^{\sqrt{\Delta\tau U}\lambda\psi_m
\tilde{n}}+O\left[(\Delta\tau)^4\right],
\label{eq:DHS}
\end{equation}
to order $(\Delta\tau)^4$. Here, $\tilde{n}= n_{1\uparrow}-n_{1\downarrow}+n_{2\uparrow}-n_{2\downarrow}
=0,\pm1,\pm2$, $\psi_{1,2}=\sqrt{2(3\mp\sqrt{6})}$ and $a_{1,2}=1\pm\sqrt{2/3}$.
All of our DQMC simulations were conducted for $U=4$ and inverse temperatures extending up to $\beta=20$.
For these parameters we used a Trotter step $\Delta\tau=0.1$, see Ref. \cite{suppmat}. In the following we present our results,
obtained by averaging over 40,000-70,000 sweeps, for systems with periodic boundary conditions and sizes of
up to $20\times20$.

\section{Results}

\subsection{Phase separation}

We begin by mapping the density as a function of the chemical potential with attention to the question
of phase separation \cite{suppmat}. The existence of phase separation in the Hubbard model has been controversial.
An early DQMC study \cite{Moreo-PS-Hubbard} found no evidence for it when $t'=0$, while subsequent
studies using the dynamical and variational cluster approximations  \cite{Maier-PS-Hubbard,VCA-PS-Hubbard}
reported its presence for both $t'>0$ and $t'=0$. To determine the presence or absence of phase separation
in the model studied here, we fix the chemical potential and follow the distribution of the hole concentration
$\delta$ throughout the Monte Carlo sampling. A bimodal distribution of $\delta$ in the thermodynamic limit
serves as an indicator for phase separation.

\begin{figure}[h!!!]
\centering
\includegraphics[width=\linewidth,clip=true]{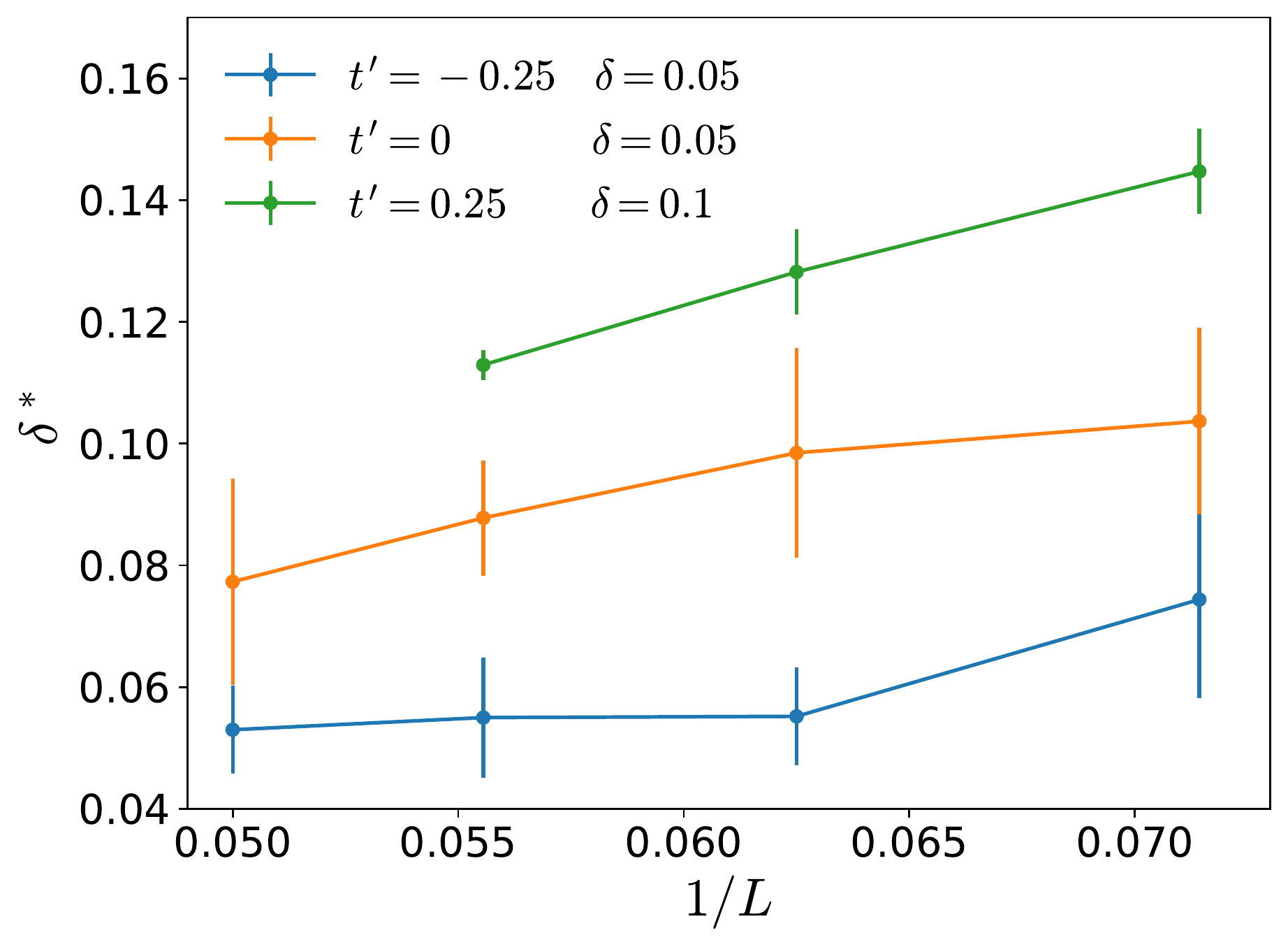}
\caption{The size dependence of $\delta^*$ in representative systems with fixed average hole concentration $\delta$.
The error bars depict the standard deviation of the distribution of hole concentrations that comprise the $\delta^*$ peak.
For the $L=20$, $t'=0.25$ system we find that all Monte Carlo configurations exhibit the same hole concentration $\delta=0.1$.}
\label{fig:ps_scalling}
\end{figure}

Our results for square $L\times L$ bilayers exhibit a bimodal distribution at low hole doping levels,
as depicted in Fig. \ref{fig:PS}. Specifically, the data in the apparent phase separated regime comprise two peaks,
one at half filling $(\delta=0)$ and another that is distributed around an average $\delta^*$.
However, when we fix the average hole concentration $\delta$ and increase the system size
we find that $\delta^*$ approaches $\delta$ and that the $\delta^*$ peak increases at the expense of the peak
at $\delta=0$. This behavior, shown in Fig. \ref{fig:ps_scalling} for representative systems with up
to $L=20$, indicates that there is no phase separation in the thermodynamic limit. In particular,
the $t'=0.25$, $\delta=0.1$ system exhibits $\delta^*=2/L$ in the range $L=14-18$ (associated with a configuration
of two filled charge stripes, as discussed below), whereas already at $L=20$ we
find a single peak at $\delta=0.1$ with no additional component at half filling.
The $t'=0$ and $t'=-0.25$ systems continue to exhibit bimodal distributions up to $L=20$, but extrapolating the
data suggests that $\delta^*\rightarrow\delta$ upon further increase of the system size.

\subsection{Stripe phases}

\begin{figure}[t!!!]
\centering
\includegraphics[width=\linewidth,clip=true]{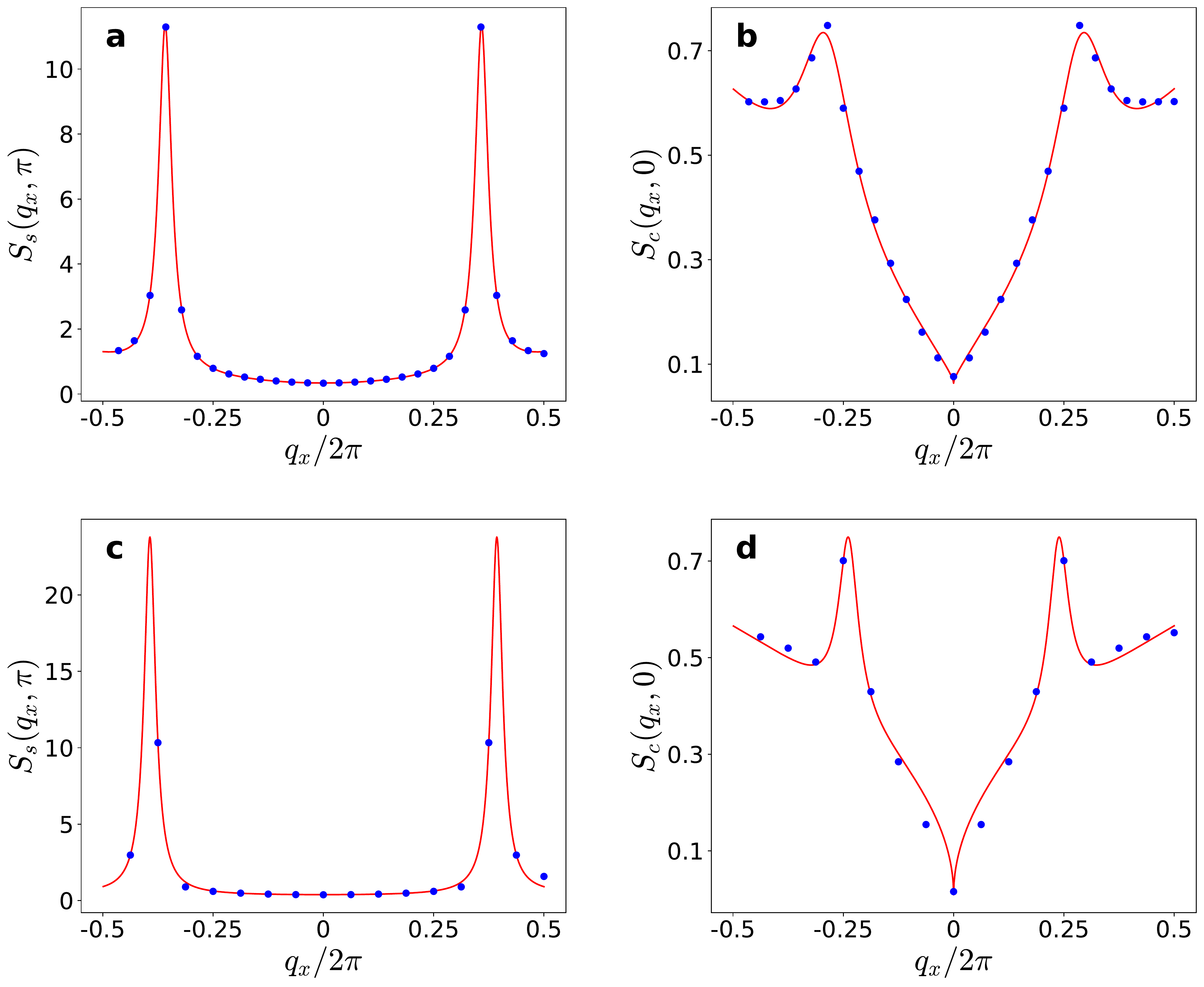}
\caption{(a) The spin structure factor $S_s(q_x,\pi)$ and (b) The charge structure factor $S_c(q_x,0)$
as a function of $q_x$ for a $28\times 4$ periodic system with $t'=-0.25$, $\delta=0.163$ and $\beta=10$.
The lines are a fit to a double Lorentzian with an added background \cite{fitcomm}. The peaks occur at $\epsilon_s=0.14$ and
$\epsilon_c=0.29$, respectively, and stay put upon lowering the temperature.
(c,d) The same quantities for a $16\times 16$ system with the same parameters at $\beta=20$. The peaks shift from their positions in the
cylindrical system to $\epsilon_s=0.11$ and $\epsilon_c=0.24$.}
\label{fig:Sc_Ss}
\end{figure}

We have already alluded to the ample numerical evidence for the existence of robust charge and
spin stripe phases in the Hubbard model, especially for $t'\leq 0$. In order to
look for similar phases in the bilayer model we have calculated the charge and spin structure factors
\begin{equation}
\label{eq:Scs}
S_{c,s}(\bq)=\frac{1}{2}\sum_{l,i}e^{-i\bq\cdot \br_i}\llangle n_{c,s}(l,\br_i) n_{c,s}(l,{\bm 0})\rrangle,
\end{equation}
with $n_{c,s}(l,\br_i)=n_{l i\uparrow}\pm n_{l i\downarrow}$. Henceforth, double angle brackets
denote connected correlation functions, i.e., $\llangle AB\rrangle=\langle AB\rangle - \langle A\rangle\langle B\rangle$.
We have found that over a wide range of parameters
$S_s$ exhibits a peak at an ordering wavevector $\bQ_s= 2\pi(0.5-\epsilon_s,0.5)$ which is typically accompanied
by a peak of $S_c$ at $\bQ_c=2\pi(\epsilon_c,0)$. For square systems we have observed similar features also along
the $y$ direction due to rotated configurations of unidirectional stripes. Representative examples are shown in
Fig. \ref{fig:Sc_Ss}. Peaks at the same momenta also occur in the charge and spin susceptibilities
\begin{equation}
\label{eq:chics}
\chi_{c,s}(\bq)=\frac{1}{2}\int_0^\beta d\tau \sum_{l,i}e^{-i\bq\cdot \br_i}\llangle n_{c,s}(l,\br_i,\tau)
n_{c,s}(l,{\bm 0},0)\rrangle.
\end{equation}
Because the $SU(2)$ spin rotation symmetry is explicitly broken by the inter-layer interaction, the spin structure
factor and the spin susceptibility differ between the $z$ and $x$-$y$ directions. We show results for their
$z$-component, defined by Eqs. (\ref{eq:Scs}) and (\ref{eq:chics}), for which the peaks are clearly visible.
Whenever peaks occur in the $S_z$ spin susceptibility they are also present
at approximately the same $\bQ_s$ in the susceptibility of the transverse spin components \cite{suppmat}.
Nevertheless, whereas the height of the former decreases by more than two orders of magnitude as one
moves from half filling to $\delta=0.3$, the latter are essentially doping-independent
and become comparable to their $z$-counterparts only at high doping levels. We do not find peaks
in the transverse spin structure factor.

\begin{figure*}
\centering
\subfigure{\includegraphics[width=0.48\textwidth]{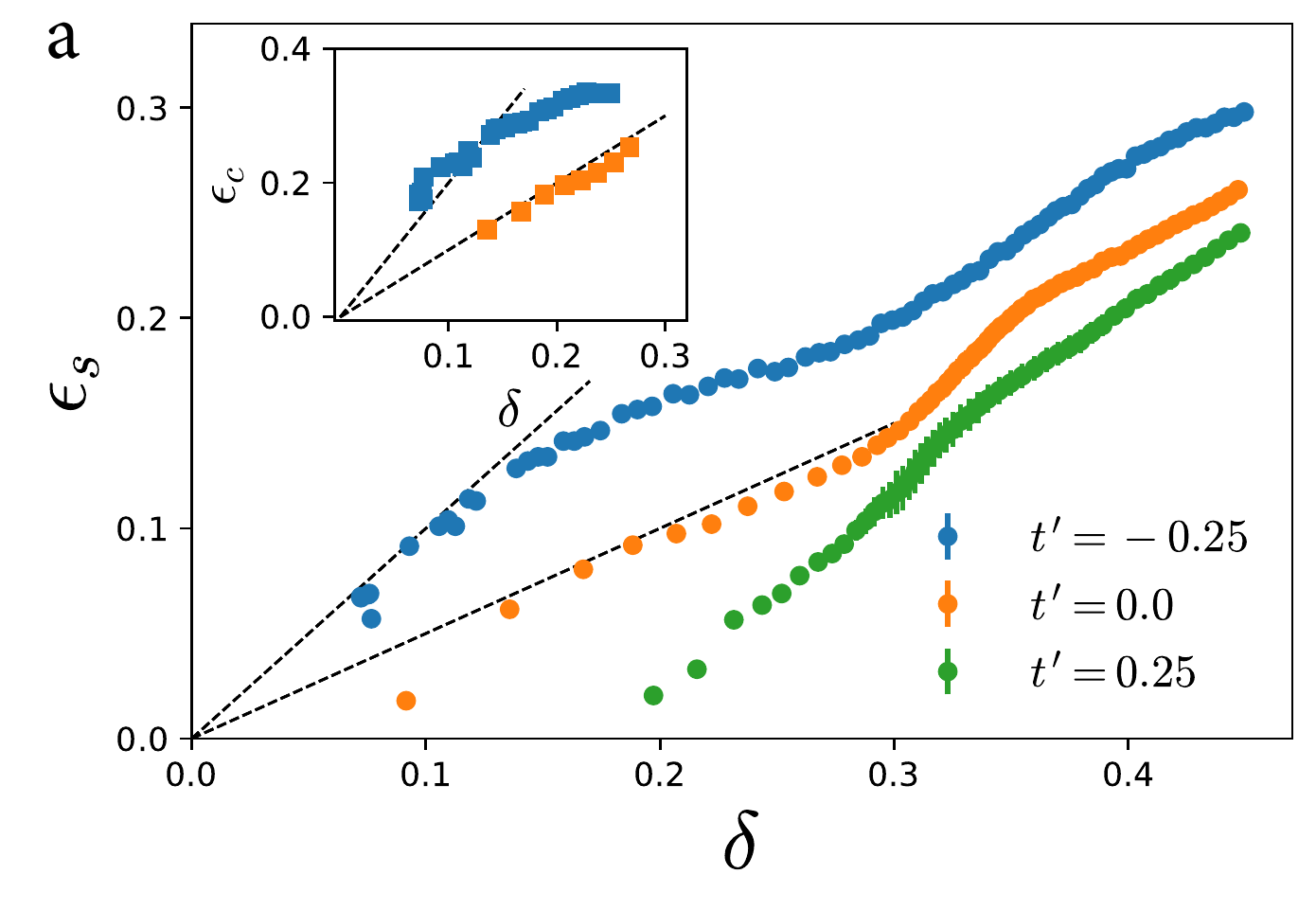}}
\subfigure{\includegraphics[width=0.48\textwidth]{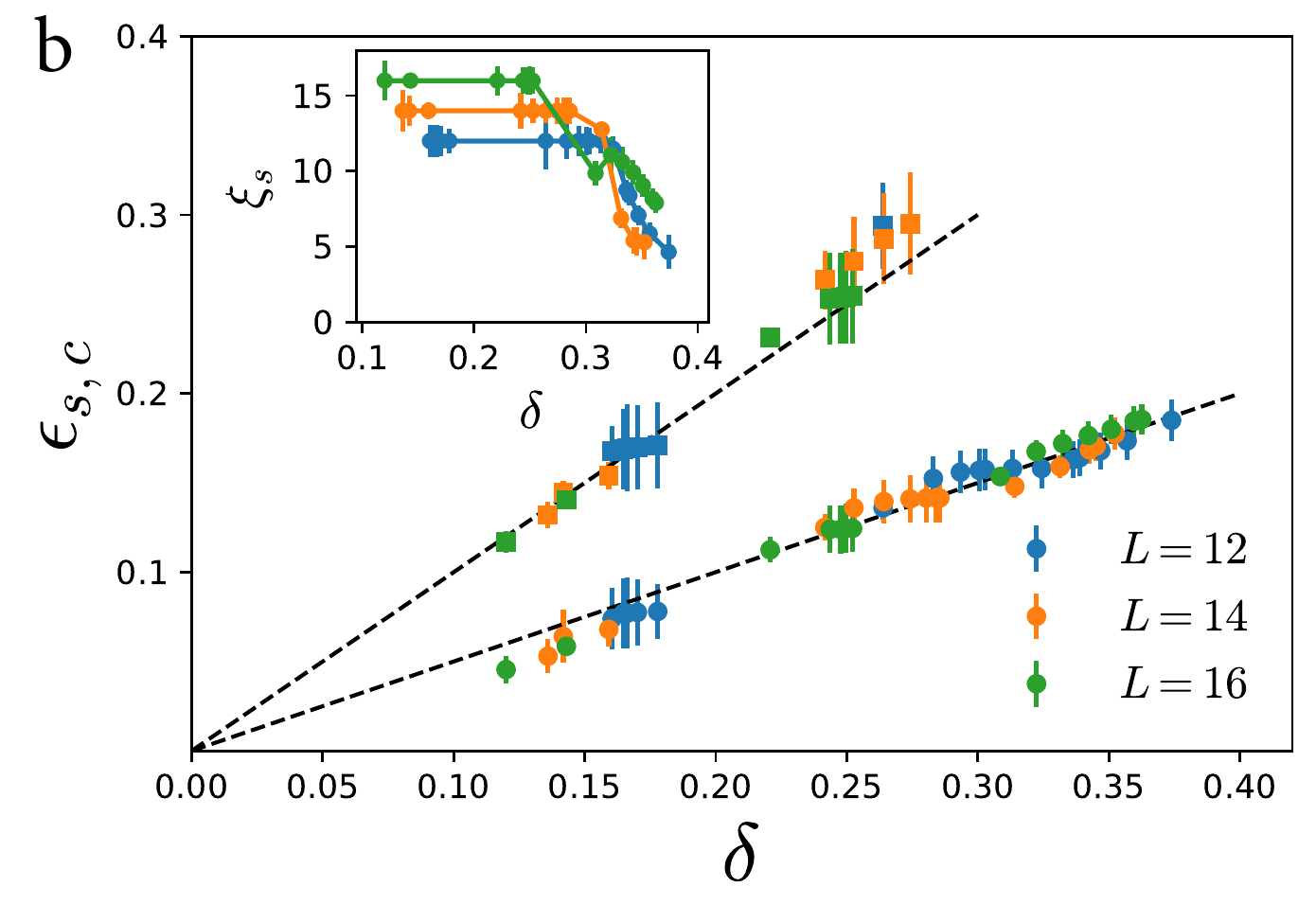}}
\subfigure{\includegraphics[width=0.48\textwidth]{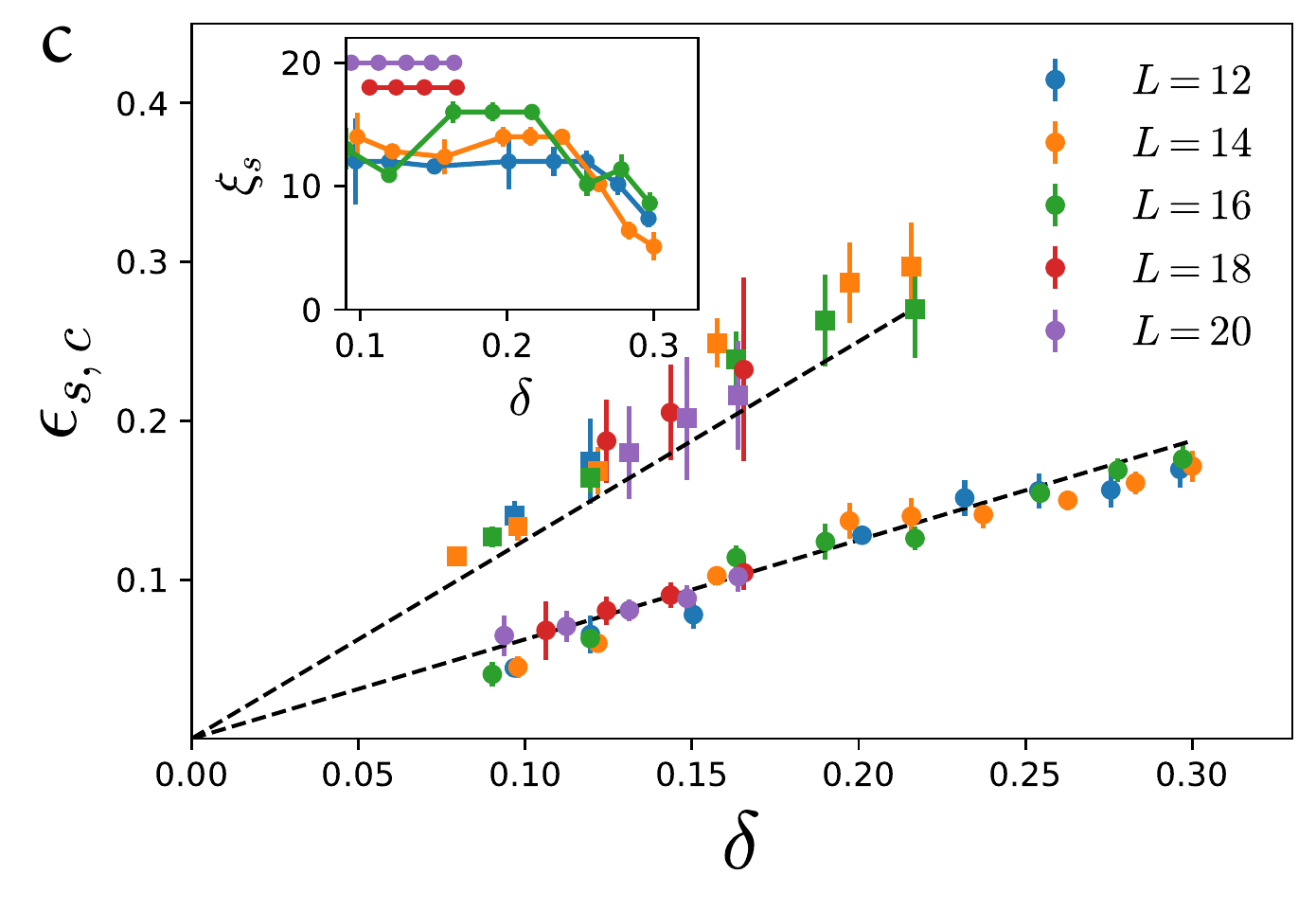}}
\subfigure{\includegraphics[width=0.48\textwidth]{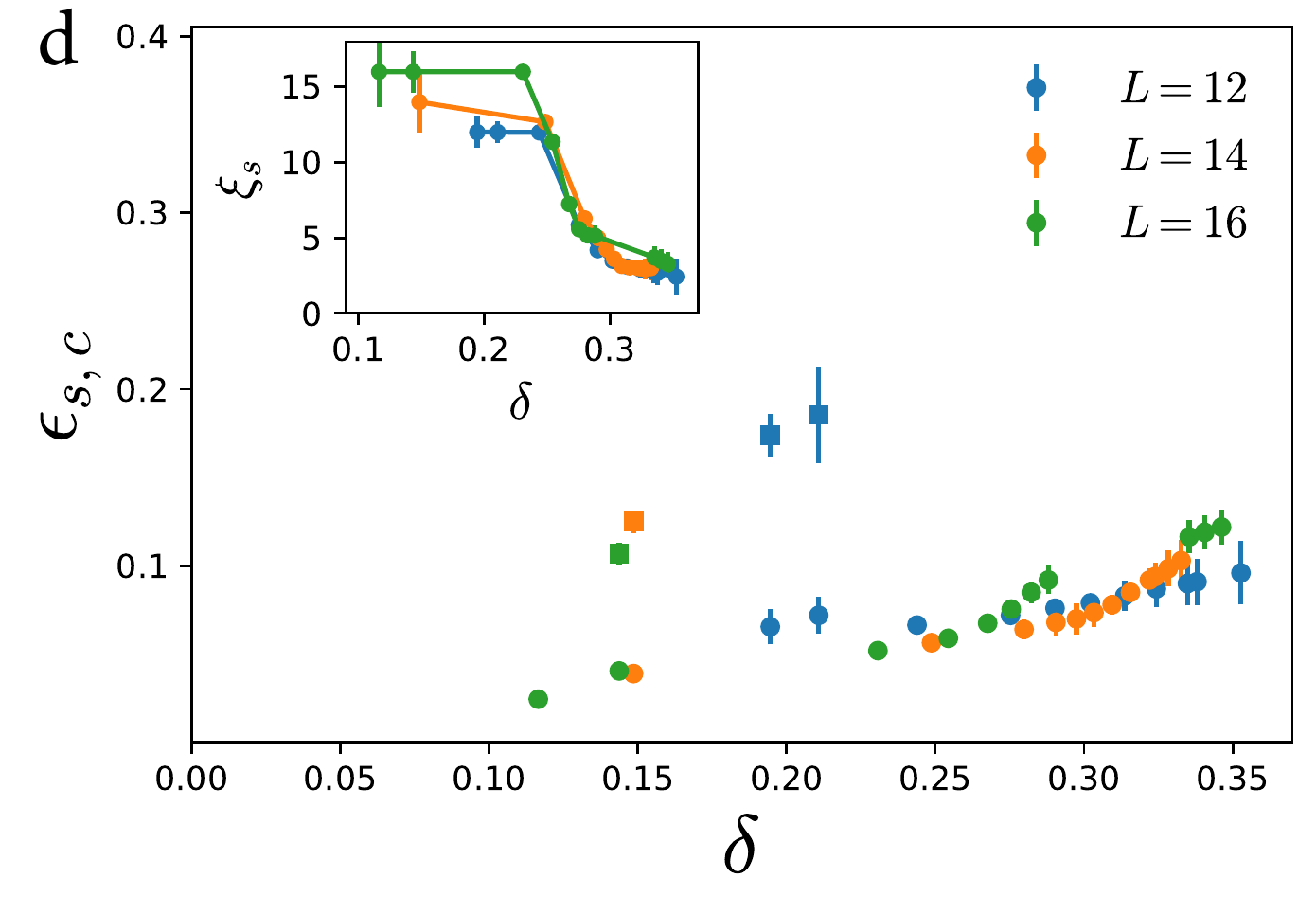}}
\caption{(a) The position of the peak in $S_s$ for a $28\times 4$ periodic bilayer as a function of hole doping
at $\beta=10$. The dashed lines correspond to $\epsilon_s=\delta/2$ and $\epsilon_s=\delta$. The inset depicts
the associated peak position in $S_c$, where here the dashed lines trace $\epsilon_c=\delta$ and
$\epsilon_c=2\delta$. (b) The position of the peak in $S_s$ (circles) and in $S_c$ (squares) for $t'=0$ periodic
square systems as a function of hole doping at $\beta=20$. The dashed lines correspond to $\epsilon_s=\delta/2$
and $\epsilon_c=\delta$. The inset depicts the correlation length of the spin stripes $\xi_s=2\pi/\Delta q$, where
$\Delta q$ is the full width at half maximum of the Lorentzian fit to the peak in $S_s$. (c,d) Similar data for
systems with $t'=-0.25$ and $t'=0.25$, respectively. The slopes of the dashed lines are 5/8 and 5/4.}
 \label{fig:combstripes}
\end{figure*}

In order to establish contact between the bilayer model and the Hubbard model, we computed the structure factors
of a quasi-one-dimensional periodic bilayer of size $28\times4$.
Fig. \ref{fig:combstripes}a depicts the positions of the peaks in $S_s$
and $S_c$ as a function of $\delta$. For $t'=0$ we observe sharp spin peaks that follow $\epsilon_s\simeq \delta/2$
within a range of doping levels that extends from 0.08 to about 0.3. Over a considerable portion of this range they
are accompanied by charge peaks at $\epsilon_c\simeq \delta$. These signatures are similar to the findings of
DMRG \cite{4leg-Hubbard-Science,4leg-Hubbard-cylinder,4leg-PRR} and DQMC studies
\cite{QMC-Hubbard-stripes,QMC-Hubbard-JPSJ} of a Hubbard system with the same geometry,
and correspond to charge stripes that host one hole per unit length and which serve as
$\pi$-phase shift domain walls for the antiferromagnetic order. We attribute the absence of stripes at small
$\delta$ to finite size effects, as one can not embed more than a single stripe within the system while
preserving the relation $\epsilon_c=\delta$. Instead, we observe in this regime alternations of the DQMC
configurations between a half filled phase and a phase with $\delta=1/L_x$, which may be associated with
a single charge stripe.

The four-leg bilayer model and the corresponding Hubbard system continue to exhibit similar stripy charge and
spin correlations also when next-nearest neighbor hopping is included. For $t'=-0.25$ and below $\delta=0.15$
we observe $\epsilon_c\simeq 2\delta$, which indicates that the charge stripes are half filled, as found
for the Hubbard cylinder \cite{4leg-Hubbard-Science}. However, the density of holes on the stripes increases
when $0.15<\delta<0.3$. In both doping ranges the period of the spin modulations is twice that of the
charge density. Here again, we associate the fact that we do not observe stripes at small $\delta$ with finite
size effects. In contrast, the absence of stripes below $\delta=0.18$ for $t'=0.25$ seems to be a true property
of the thermodynamic limit of the model, at least in the temperature range that we have considered. This is
consistent with the DQMC results for the Hubbard system \cite{QMC-Hubbard-stripes}. We note that at
doping levels above $\delta=0.3$ and for all values of $t'$, the model exhibits a phase with short-ranged stripe
correlations that are accompanied by a change in the dependence of $\epsilon_s$ on $\delta$,
see Fig. \ref{fig:combstripes}a. This range of parameters has not been investigated in the context of the
Hubbard cylinder and it would be interesting to close this gap in order to see if the similarities between
the models continue to hold true for high doping levels.

Having mapped out the behavior of the quasi-one-dimensional system we proceed to discuss the
stripe characteristics of more two-dimensional geometries, which are not amenable to DMRG calculations.
Accordingly, we have computed $S_s$ and $S_c$ of $L\times L$ periodic bilayers, with $L=12 - 20$.
The results for $t'=0$, which are depicted in Fig. \ref{fig:combstripes}b, show the same linear doping
dependence $\epsilon_s=\delta/2$ and $\epsilon_c=\delta$ as in the four-leg bilayer. However, in the square
systems the linear dependence does not change across the transition from the region where
the correlation length of the spin stripes, $\xi_s$, exceeds the system size to the regime where $\xi_s<L$.
Fig. \ref{fig:combstripes}c demonstrates that changing the hopping amplitude to $t'=-0.25$ has little effect
on the doping range that supports stripes and on the stripes correlation length. At the same time, the slope
of $\epsilon_c(\delta)$ increases, thereby implying that the number of holes per unit length of a charge
stripe reduces from 1 for $t'=0$ to 2/3-4/5 when $t'=-0.25$. This observation bear resemblance to the findings
of a recent DMRG study of a six-leg Hubbard cylinder \cite{6legHubbard}. Finally, the square $t'=0.25$ systems
show signatures of fractionally filled charge stripes that we did not detect in the four-leg torus.
Furthermore, spin stripes appear at lower doping levels in the square systems than in the four-leg bilayer, see
Fig. \ref{fig:combstripes}d. In fact, given the limited range of system sizes available to us, we are unable
to exclude the existence of spin stripes at even lower values of $\delta$ as $L$ is further increased.

\subsection{Superconductivity}

The question of whether the two-dimensional repulsive Hubbard model exhibits superconductivity at a temperature
scale that is relevant to the cuprate superconductors has been the focus of extensive research over the years.
The current evidence points to a negative answer when $t'=0$ \cite{Hubbard-absence}, and arguably also for
$t'<0$ \cite{6legHubbard}. The situation is somewhat more promising for $t'>0$, where power-law
$d$-SC correlations are detected, albeit with a faster decay than the CDW correlations \cite{6legHubbard}.
Hence, it is interesting to look for signs of superconductivity in the bilayer model, with emphasis on
$d$-SC, which is expected to be the dominant channel in the presence of repulsive interactions.

To this end, we have calculated the intra-layer $d$-wave pair-field susceptibility
\begin{equation}
\chi_d=\frac{1}{2}\int_0^{\beta}d\tau \sum_{l=1,2}\sum_{i}\langle \Delta _{d}(l,\br_i,\tau)
\Delta_{d}^{\dagger}(l,{\bm 0},0)\rangle,
\label{eq:chi_d}
\end{equation}
where $\Delta_{d}(l,\br_i)=\frac{1}{4}\sum_{\alpha=\pm\hat{x},\pm\hat{y}}\eta_{\alpha}
(c_{li\uparrow}c_{li+\alpha\downarrow}-c_{li\downarrow}c_{li+\alpha\uparrow})$
with $\eta_\alpha=1$ for $\alpha=\pm\hat{x}$ and $\eta_{\alpha}=-1$ for $\alpha=\pm\hat{y}$.
To reveal the effects of interactions on the superconducting properties we have also evaluated
the particle-particle interaction vertex
\begin{equation}
\Gamma=\frac{1}{\chi_d}-\frac{1}{\bar{\chi}_d},
\end{equation}
where $\bar{\chi}_d$ is the uncorrelated $d$-wave pair-field susceptibility \cite{White-vertex}.
Onset of superconducting quasi-long-range order in the two-dimensional thermodynamic limit
manifests itself by an increase of $\chi_d$ with decreasing temperature. In a finite system of
linear size $L$ the temperature dependence should also exhibit BKT finite-size scaling of the form
$\chi_d=L^{7/4}f(L/\xi)$, where $\xi$ is the BKT correlation length, see e.g. Refs. \cite{chi-scaling,Paiva2}.
In particular, one expects $\chi_d$ to saturate at low temperatures at a value that increases with $L$.
Concomitantly, if the interactions indeed drive the system towards a superconducting instability then
the product $\Gamma \bar{\chi}_d$ should approach -1 at the critical temperature.

\begin{figure}[t!!!]
\centering
\includegraphics[width=\linewidth,clip=true]{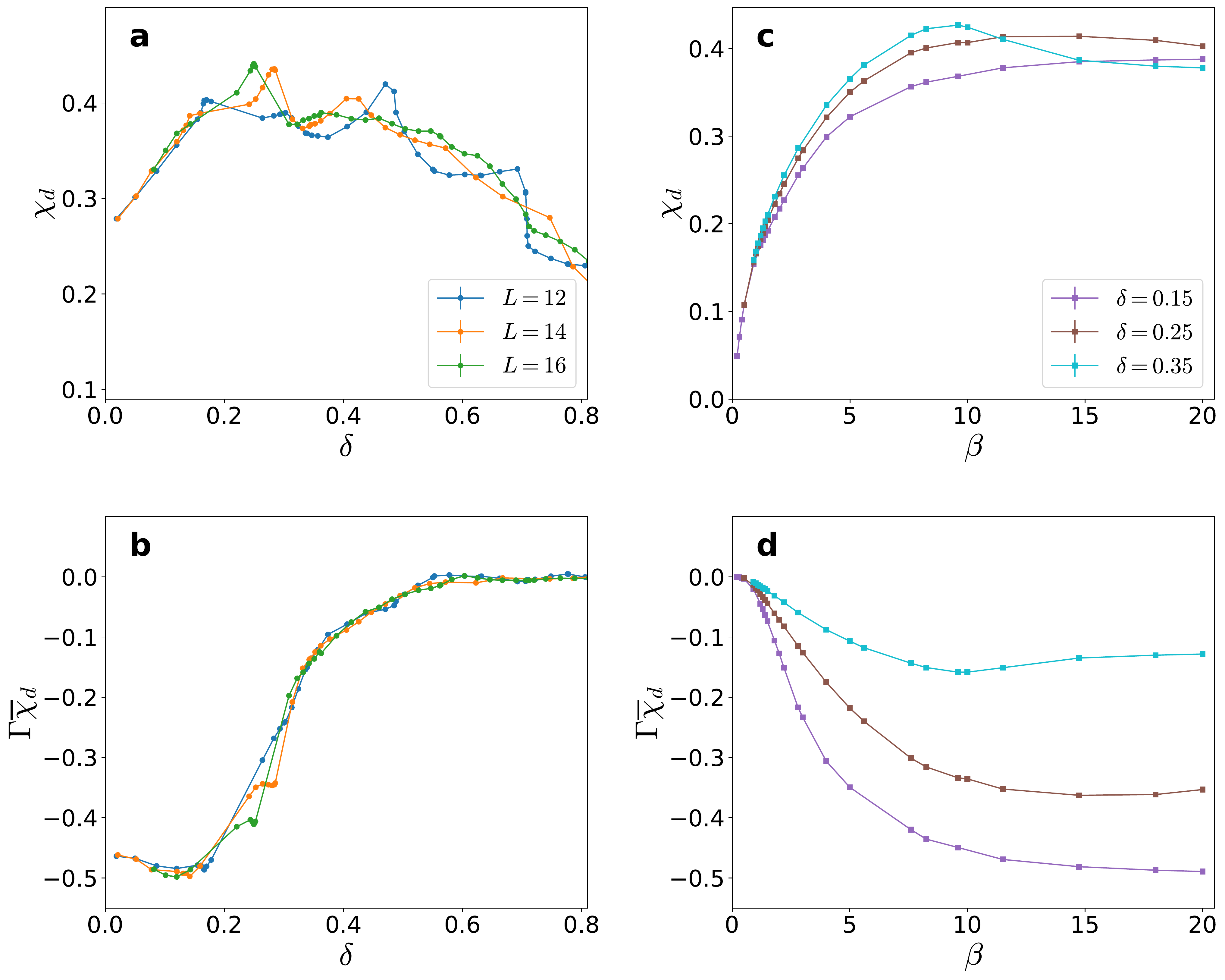}
 \caption{(a) The intra-layer $d$-wave pairing susceptibility as a function of hole doping for $L\times L$
 periodic systems with $t'=0$ at $T=0.05$. (b) The $d$-wave superconducting vertex times the
 uncorrelated pairing susceptibility of the same systems. (c,d) The same quantities as a function of inverse
 temperature $\beta$ for an $L=14$ bilayer at the specified hole doping levels.}
\label{fig:sc_pd_0}
\end{figure}
\begin{figure}[t!!!]
\centering
  \includegraphics[width=\linewidth,clip=true]{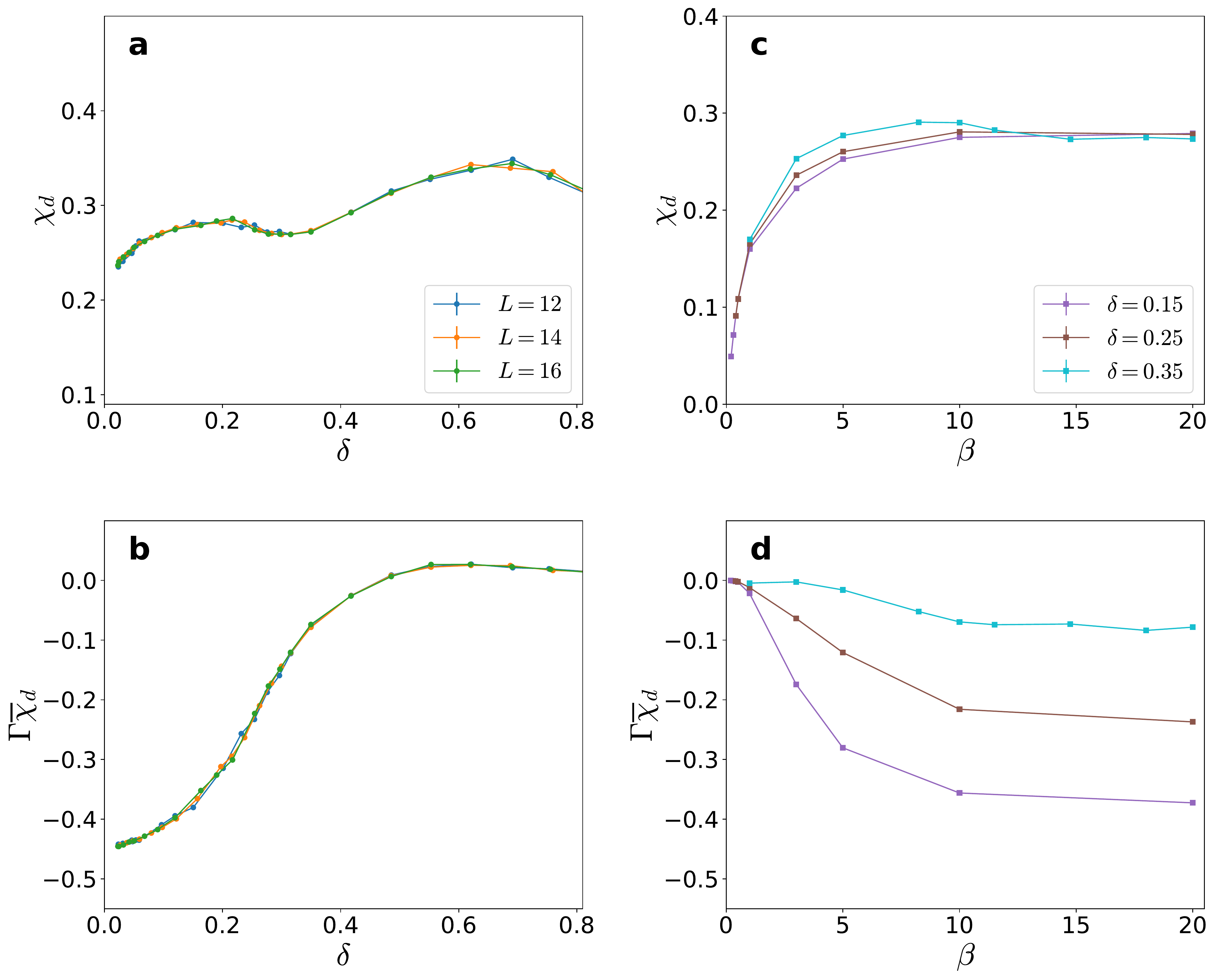}
 \caption{Same as Fig. \ref{fig:sc_pd_0} but for $t'=-0.25$}
  \label{fig:sc_pd_-0.25}
\end{figure}

\begin{figure}[t!!!]
\centering
  \includegraphics[width=\linewidth,clip=true]{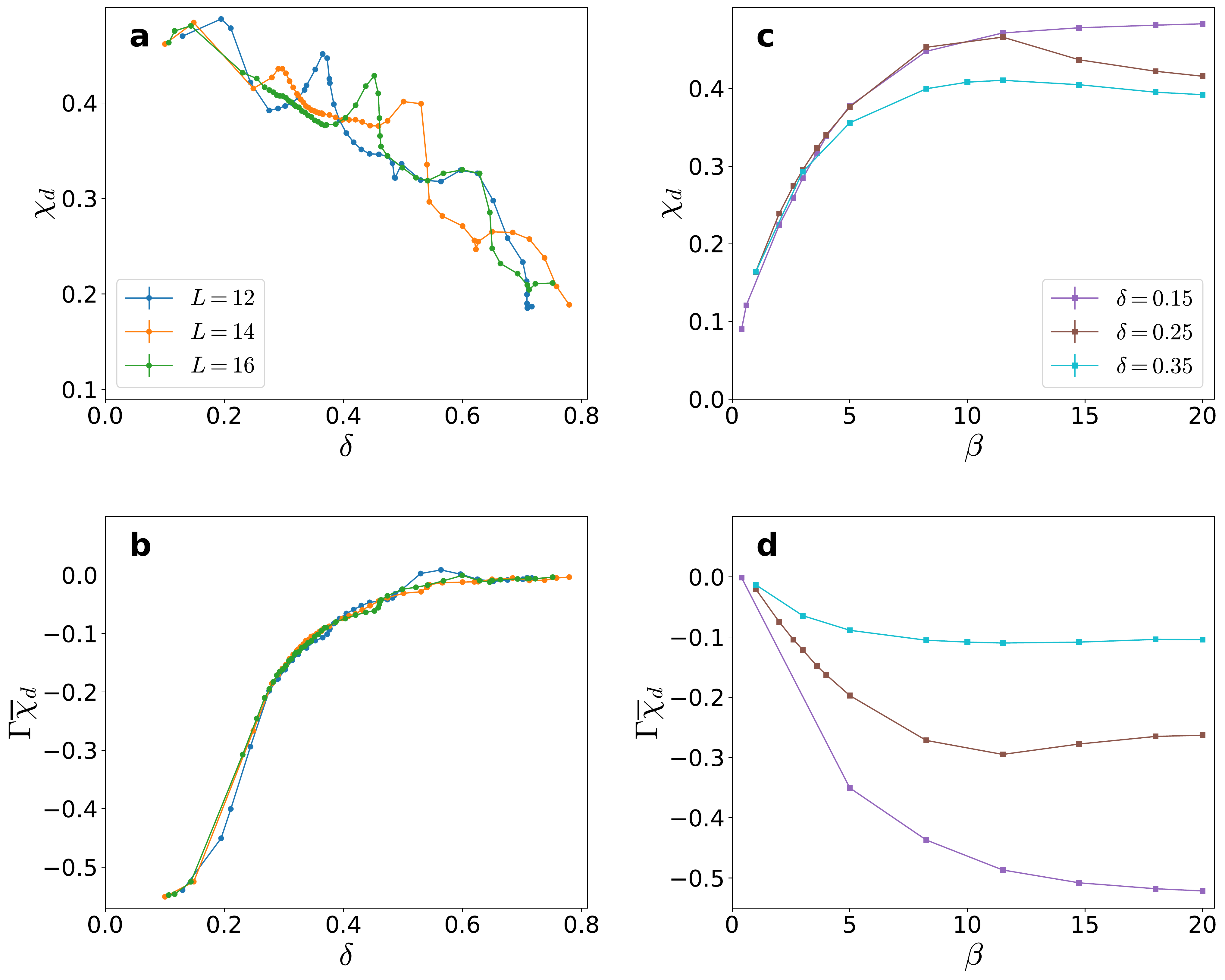}
 \caption{Same as Fig. \ref{fig:sc_pd_0} but for  $t'=0.25$}
  \label{fig:sc_pd_0.25}
\end{figure}

We find none of the above signatures in the data for $t'=0$ bilayers, as presented in Fig. \ref{fig:sc_pd_0}.
In particular, it is clear that both $\chi_d$ and $\Gamma \bar{\chi}_d$ are already saturated at the lowest
temperature, $T=0.05$, that we have considered. However, both quantities show no significant size dependence
at this temperature, with $\chi_d$ exhibiting some fluctuations as a function of $\delta$, which we attribute
to finite size effects. Furthermore, while $\chi_d$ reaches a maximum around $\delta=0.25$, $\Gamma \bar{\chi}_d$
attains its minimal value of about -0.5 near $\delta=0.1$. The lack of correlation between the doping dependence
of the two functions is further evidence that the $t'=0$ bilayer shows no signs of a $d$-SC instability.
Fig. \ref{fig:sc_pd_-0.25} shows that a similar behavior is found for $t'=-0.25$. If at all, the indications
for $d$-SC are weaker, in the sense that $\chi_d$ is maximal at $\delta=0.7$ whereas the minimum of
$\Gamma \bar{\chi}_d$ occurs near half filling and is slightly higher than its $t'=0$ value. These findings
conform well with the results of a DQMC study of a $t'=-0.25$ single Hubbard layer, albeit at higher temperatures
\cite{QMC-Hubbard-JPSJ}. Finally, the systems with $t'=0.25$ exhibit the most favorable hints for the existence
of some $d$-SC tendencies. Specifically, $\chi_d$ and $\Gamma \bar{\chi}_d$ show simultaneous maximal response
at low doping levels below $\delta=0.2$, that is also the largest among the bilayers that we have studied, see
Fig. \ref{fig:sc_pd_0.25}. Nevertheless, the fact that the response is still far from the instability threshold
and does not show the expected finite-size scaling leads us to conclude that $d$-SC does not materialize in the
bilayer model, at least for the parameters used by us.

One may object to the sweeping nature of the last statement as we have only referred to signs of uniform $d$-SC
order. Indeed, there have been suggestions that the cuprate superconductors and perhaps some theoretical models
may harbor the more elusive pair-density wave (PDW) state that is associated with a spatially oscillating
superconducting order parameter of zero mean \cite{PDW-review}. On the theoretical side, the stabilization
of a PDW in interacting fermionic models has been proved difficult, with the best evidence for PDW correlations
emerging from DMRG studies of a one-dimensional Kondo-Heisenberg chain \cite{Berg-KH} and of the strong-coupling
limit of a Holstein-Hubbard cylinder \cite{Kivelson-PDW-Holstein}. To address the possibility of the existence
of $d$-wave pairing with a non-zero center-of-mass momentum we have calculated the Fourier transform of the
pair-field susceptibility in Eq. (\ref{eq:chi_d}). However, our results show a single and robust peak of
$\chi_d(\bq)$ at $\bq=0$ with no evidence for a PDW.

\begin{figure}[t!!!]
\centering
\includegraphics[width=\linewidth,clip=true]{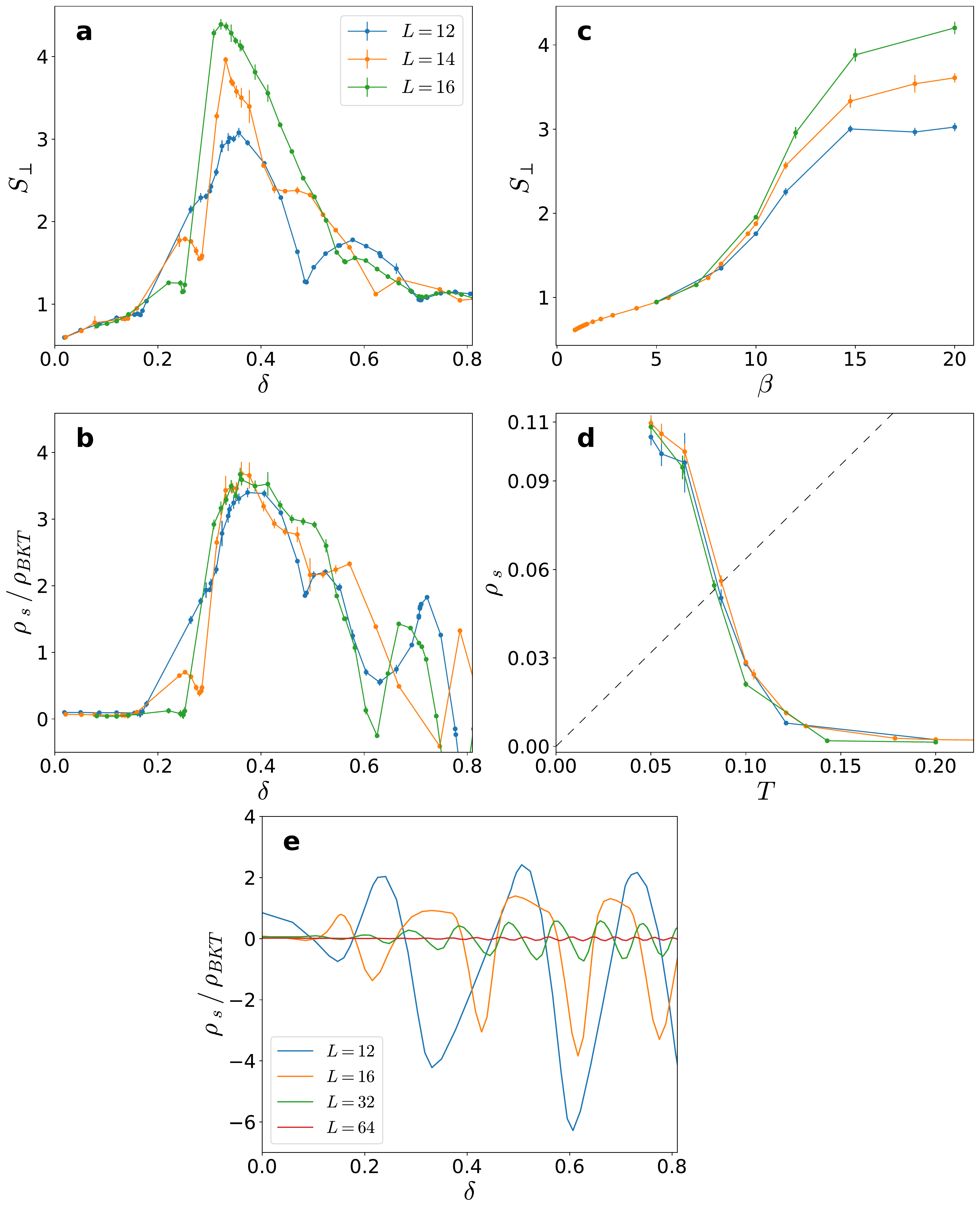}
\caption{(a) The equal-time inter-layer pair correlations as a function of doping for $L\times L$
periodic systems with $t'=0$ at $T=0.05$. (b) The superfluid stiffness of the
systems, normalized by $\rho_{BKT}=2T/\pi=0.1/\pi$. (c,d) The pair correlations and the
superfluid stiffness as a function of temperature for $\delta=0.35$. The dashed line depicts $2T/\pi$.
(e) The superfluid stiffness of non-interacting ($U$=0) bilayers at $T=0.05$.}
\label{fig:sc_P_perp_0}
\end{figure}

Despite not finding a pairing instability in the intra-layer $d$-wave channel we have not exhausted
the search for superconductivity. Indeed, a more universal indicator of superconductivity is the superfluid
stiffness, calculated from the response to a vector potential that couples identically to the two layers
via \cite{Stiffness-def}
\begin{equation}
\rho_s=\frac{1}{4}\left[\Lambda_{xx}(q_x\rightarrow 0,q_y=0)-\Lambda_{xx}(q_x=0,q_y\rightarrow 0)\right],
\label{eq:rho_s}
\end{equation}
where
\begin{equation}
\Lambda_{xx}(\bq)=\frac{1}{2L^2}\sum_{l,l'=1,2}\int_0^{\beta}d\tau \langle j_x(\bq,l,\tau)j_x(-\bq,l',0)\rangle.
\end{equation}
Here,
$j_x(\bq,l)=-i\sum_{j,\sigma}\{c^\dagger_{l,j,\sigma}[tc_{l,j+\hat{x},\sigma}+t'(c_{l,j+\hat{x}+\hat{y},\sigma}
+c_{l,j+\hat{x}-\hat{y},\sigma})]-{\rm H.c.}\}e^{-i\bq\cdot \br_j}$, is the Fourier transform of the
current density operator in the $x$ direction. In the finite $L\times L$ bilayers that we simulate we
obtain the limit $q\rightarrow 0$ in Eq. (\ref{eq:rho_s}) by extrapolating $\Lambda_{xx}$ using its values
at $q=2\pi/L$ and $q=4\pi/L$. A typical temperature dependence of $\rho_s$ is depicted in Fig. \ref{fig:sc_P_perp_0}d
for $t'=0$ systems at $\delta=0.35$. Clearly, $\rho_s$ at this doping level shows little size dependence and obeys
the criterion for the BKT transition $\rho_s(T_{BKT})=(2/\pi)T_{BKT}$ at a critical temperature $T_{BKT}\simeq 0.08$.
Fig. \ref{fig:sc_P_perp_0}b shows that the bilayer undergoes a BKT transition to a superconducting state over
an extended range of hole doping. The figure depicts the ratio $\rho_s(T)/(2T/\pi)$ at $T=0.05$, such that
any point for which the ratio is larger than one corresponds to a system that exhibits a transition at
a temperature $T>0.05$. The results also demonstrate the strong finite-size effects in $\rho_s$ at high doping levels,
which cause the stiffness to oscillate between positive and negative values. Such a behavior reflects changes in
the Fermi surface and is inherited from the non-interacting limit, see Fig. \ref{fig:sc_P_perp_0}e. It is
noticeable whenever the interaction effects are diminished, as is the case for large $\delta$.
This issue can be mitigated by introducing a weak uniform magnetic field to the model \cite{Assaad-field},
but we forewent the modification since the problem is significant only in a region where the qualitative
behavior is already clear.


\begin{figure}[t!!!]
\centering
\includegraphics[width=\linewidth,clip=true]{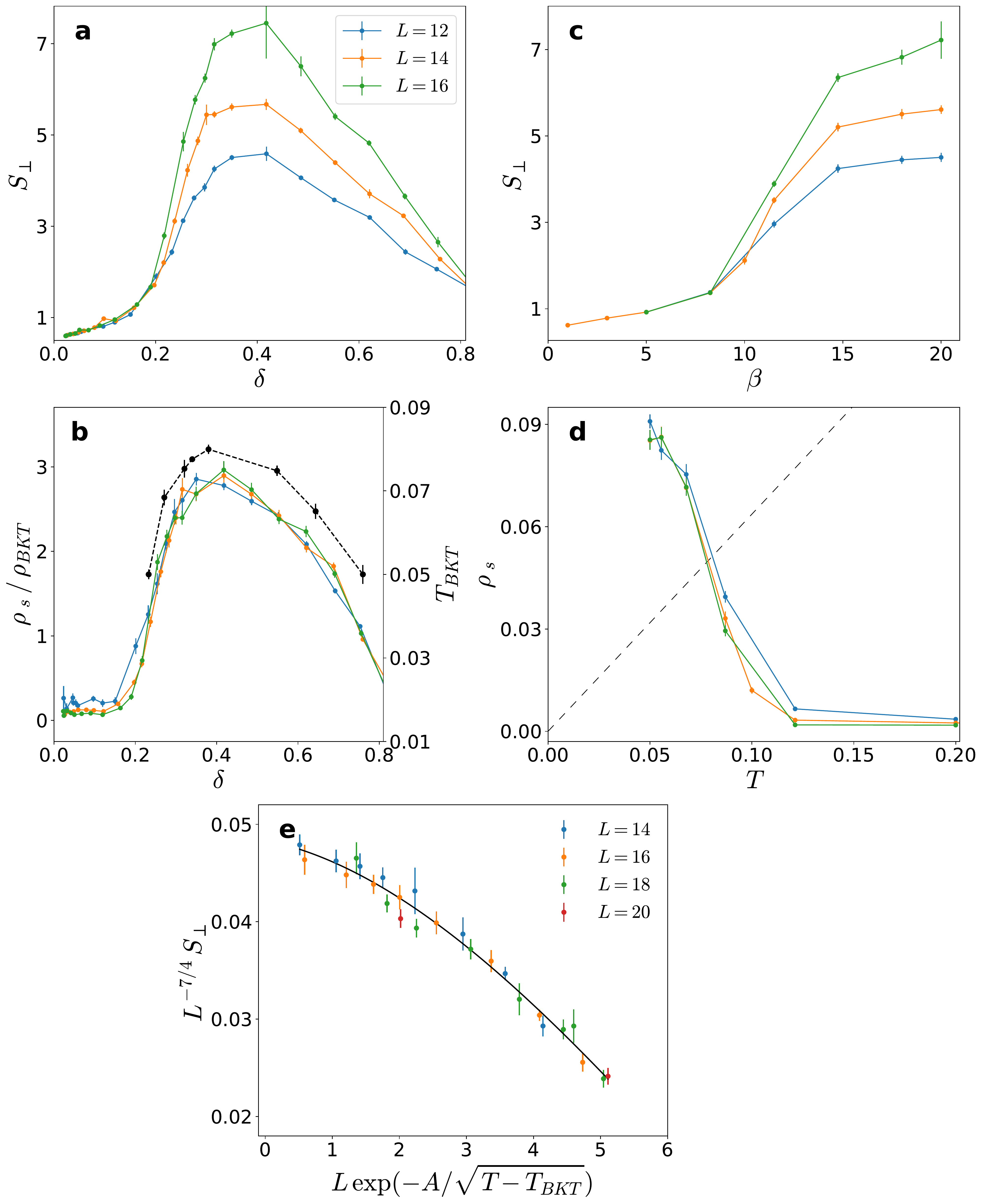}
\caption{Same as Figure \ref{fig:sc_P_perp_0} but for $t'=-0.25$. The black curve in (b) depicts the
BKT transition temperature, deduced from the BKT criterion, as a function of doping. Panel (e) depicts
the collapse of the $S_\perp$ data for a system with $\delta=0.35$
using the expected BKT scaling. The line is a guide to the eye.}
\label{fig:sc_P_perp_-025}
\end{figure}

A question remains as to the nature of the superconducting state. The presence of the ferromagnetic interaction
between the layers makes inter-layer spin polarized pairing a natural candidate for the instability channel. We
have corroborated this hypothesis by calculating the equal-time inter-layer pair correlations
\begin{equation}
S_{\perp}=\frac{1}{2}\sum_{i\sigma}\langle\Delta_{\perp \sigma}(\br_i)\Delta_{\perp \sigma}^{\dagger}({\bm 0})\rangle,
\end{equation}
where $\Delta_{\perp\sigma}(\br_i)=c_{1i\sigma}c_{2i\sigma}$. Fig. \ref{fig:sc_P_perp_0}c shows that the
temperature dependence of $S_\perp$ begins to develop size dependence slightly above $T_{BKT}$ and saturates at
low temperatures to a value that grows with $L$. These signatures strongly support the identification of $T_{BKT}$
with the onset of quasi-long-range $\Delta_\perp$ correlations. Furthermore, the doping dependence of the
low-temperature $S_\perp$ follows that of $\rho_s$ and exhibits size dependence within the range
of doping levels where the system is below its $T_{BKT}$ according to the BKT criterion,
see Fig. \ref{fig:sc_P_perp_0}a. We note that $S_\perp$ of the $t'=0$ bilayers attains its maximum around
$\delta=0.35$. This fact may be tied to the decline of the intra-layer $d$-SC susceptibility of the
$\delta=0.35$ system at temperatures below its $T_{BKT}$, as seen in Fig. \ref{fig:sc_pd_0}c.

\begin{figure}[t!!!]
\centering
  \includegraphics[width=\linewidth,clip=true]{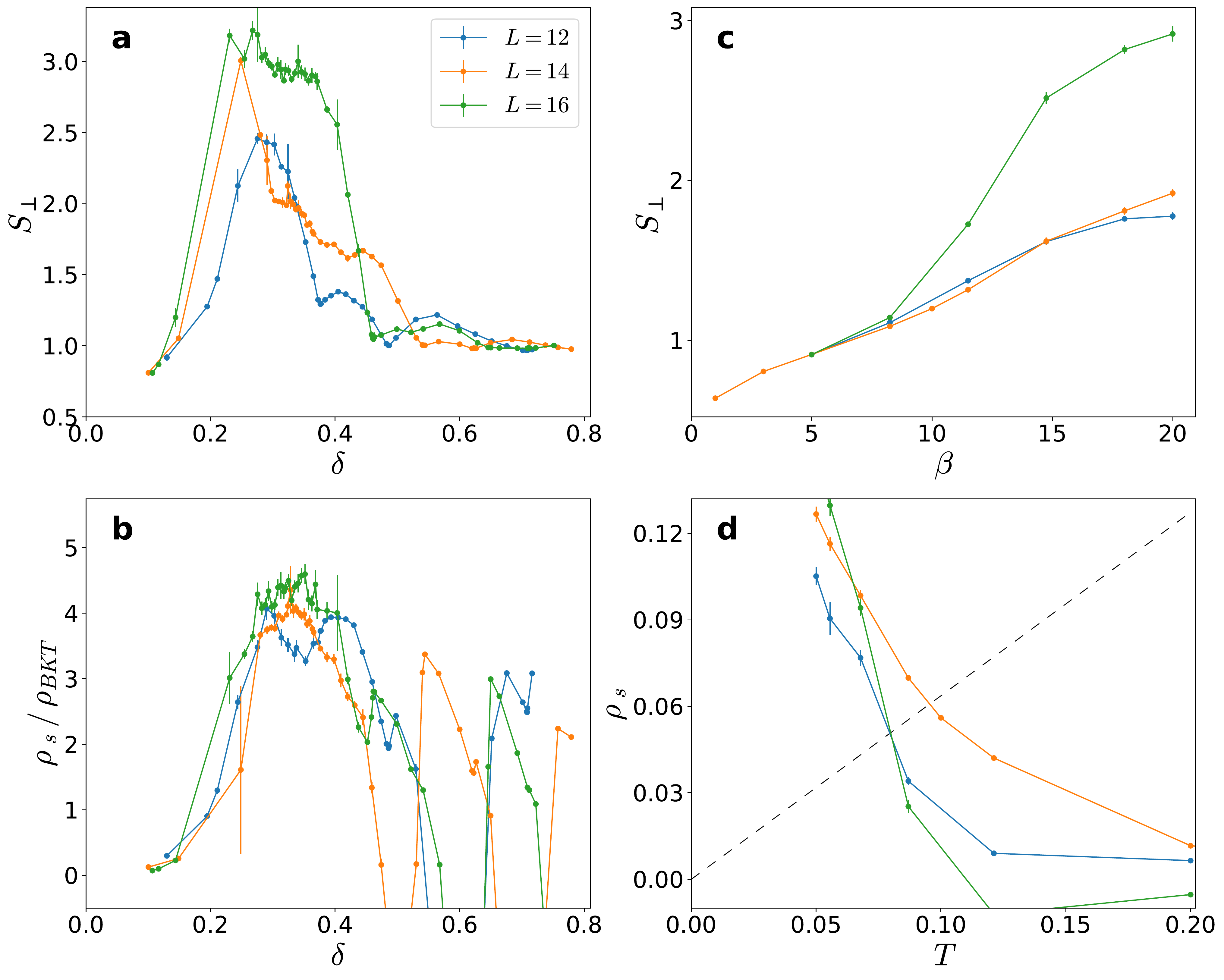}
 \caption{Same as Figure \ref{fig:sc_P_perp_0} but for $t'=0.25$}
  \label{fig:sc_P_perp_025}
\end{figure}

Both $\rho_s$ and $S_\perp$ continue to exhibit similar trends in the presence of non-zero $t'$.
Fig. \ref{fig:sc_P_perp_-025} shows that the $t'=-0.25$ bilayer sustains a superconducting phase over
a wider range of doping levels as compared to the $t'=0$ bilayer. The figure also contains
results for the doping dependence of $T_{BKT}$, as deduced from the BKT criterion,
and demonstrates that it follows the behavior of the low-temperature superfluid stiffness.
Namely, the two trace a dome as a function of $\delta$, achieving a maximum around $\delta=0.4$.
Further evidence in favor of the onset of quasi-long range order comes form applying the BKT
scaling ansatz $S_\perp(L,T)=L^{7/4}f[L/\xi(T)]$, where $\xi(T)\sim \exp[A/(T-T_{BKT})^{1/2}]$,
as $T\rightarrow T_{BKT}$ from above. Fig. \ref{fig:sc_P_perp_-025}(e) shows the scaling, where
$A=0.17$ and $T_{BKT}=0.72$ yield the best data collapse.
Fig. \ref{fig:sc_P_perp_025} shows that
for $t'=0.25$ the leading edge of the superconducting dome shifts to lower values of doping,
somewhat below $\delta=0.2$. For reasons that are not clear to us the fluctuations associated
with finite size effects are much reduced in the results for the $t'=-0.25$ bilayers, while they
are enhanced in the $t'=0.25$ systems.

\begin{figure}[t!!!]
\centering
\includegraphics[width=\linewidth,clip=true]{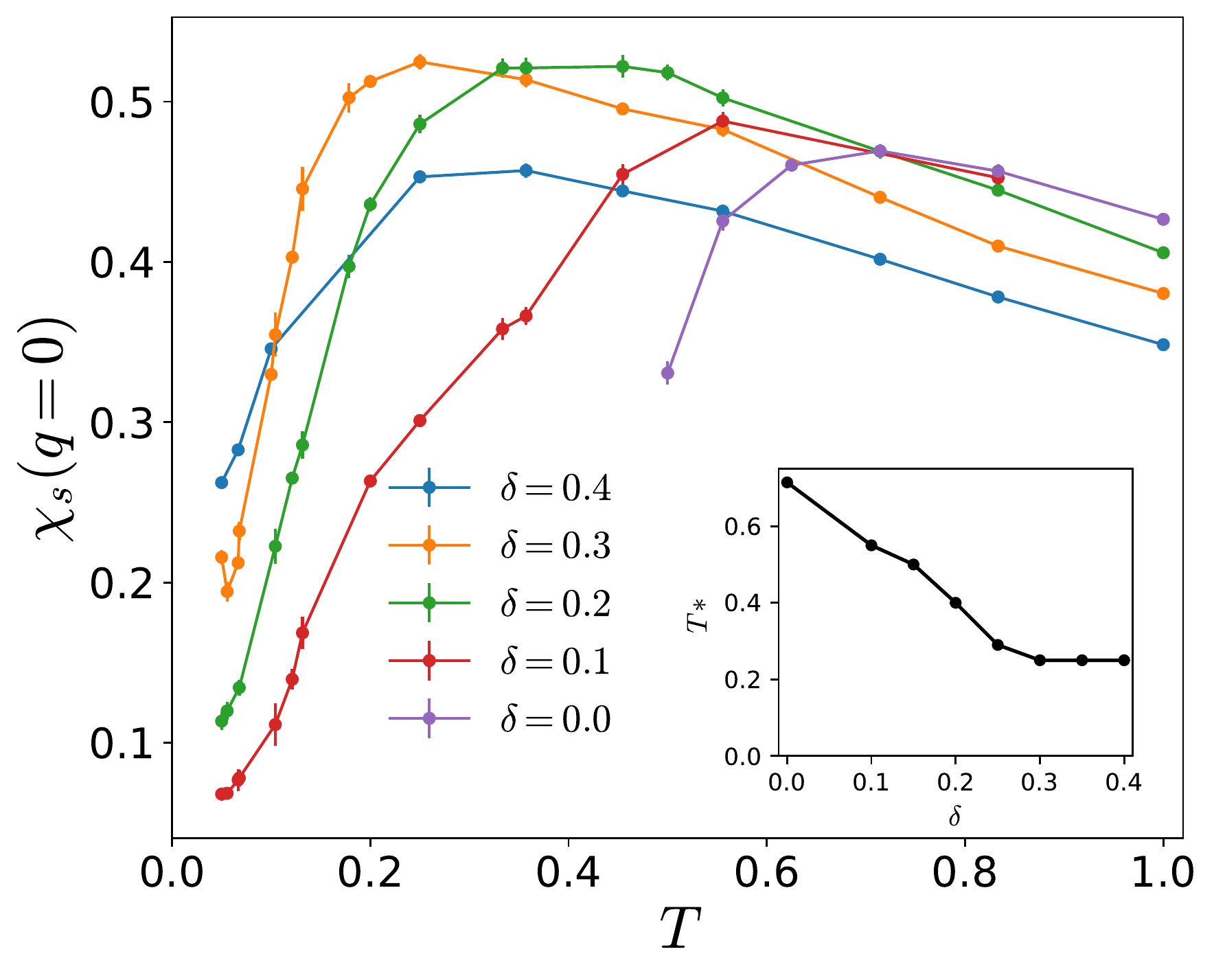}
\caption{The $q=0$ spin susceptibility (Knight shift) versus temperature for various doping levels of the
$14\times 14$ $t'=0$ system. The curves display a maximum at a temperature denoted by $T^*$. The inset
depicts $T^*$ as a function of doping.}
\label{fig:Knight_shift}
\end{figure}

We end this section by considering possible correlations between the superconducting properties
of the model and its uniform magnetic susceptibility along the $z$ direction (the Knight shift).
The temperature dependence of the Knight shift is presented in Fig. \ref{fig:Knight_shift} for a $t'=0$
system at several doping levels. We find that it exhibits a peak at a temperature $T^*$ that reduces
with increasing hole doping until $\delta\approx 0.3$, where it levels off. In the context of the
cuprates such a peak is used to define a crossover scale that is associated with the opening of
a pseudogap. As far as the model is concerned, there seems to be no clear correspondence between
$T^*$ and the behavior of the $d$-SC signatures. From Fig. \ref{fig:sc_pd_0} it is evident that the
low temperature $\Gamma \bar{\chi}_d$ does not change within the range of doping levels in which $T^*$
changes by more than a factor of 3, while $\chi_d$ increases over the same range. Both quantities
appear to saturate at a similar temperature that shows no considerable dependence on doping, and hence
on $T^*$. On the other hand, the inter-layer superconductivity onsets at a doping level which coincides
with the point where $T^*$ becomes $\delta$-independent, see Figs. \ref{fig:sc_P_perp_0} and
\ref{fig:Knight_shift}. However, the causal relation between the two phenomena is unclear.
Regardless of its relevance to superconductivity, the appearance of the peak and the general behaviour
of $T^*$ are in agreement with DQMC results obtained for the Hubbard model at more elevated temperatures
\cite{QMC-Hubbard-JPSJ}. This fact reinforces the conclusion that despite the difference in their
symmetries the bilayer model and the Hubbard model display many common physical properties.

\section{Summary and outlook}

What are the low-temperature properties of the repulsive two-dimensional Hubbard model at the intermediate
coupling regime $U\sim t$? By far, the most adopted approach towards answering this question has been to
apply DMRG to $W\times L$ systems, then try to extract the $L\rightarrow\infty$ behavior for fixed $W$
and finally look for $W$-independent characteristics within the limited range of computationally manageable
widths. Less frequent attempts involved studying the model on a more two-dimensional geometry (typically square)
using DQMC. However, this method suffers from the sign problem that constrains its application
to relatively high temperatures, hence complicating the comparison with the DMRG results for the ground state.
In the present work we have offered yet another route that applies sign-problem-free DQMC to a bilayer model
which contains the Hubbard interaction on each layer, at the expense of introducing a ferromagnetic attractive
inter-layer coupling that breaks the global $SU(2)$ spin rotation symmetry. Our primary goal was to assess
the degree by which the deformed model captures known behaviors of the Hubbard model or exhibits qualitatively
new features.

We have found that in a similar fashion to DMRG and DQMC results for Hubbard systems, the most robust tendency
of the bilayer model is to develop stripy modulations in its charge and spin densities. The agreement extends
to specific properties of the observed stripe phases. Namely, the charge stripes are filled for $t'=0$,
exhibit fractional filling that is larger than 1/2 when $t'=-0.25$, and are suppressed, especially
at low doping levels, for $t'=0.25$. The lack of spin rotation symmetry renders the $z$ and $x$-$y$ spin
responses of the model inequivalent. Nevertheless, the $z$ component of the spin density shows very
clear stripes, which become short-range correlated and appear without accompanying charge stripes
at high doping. Whenever spin and charge stripes coexist the period of the first is approximately
twice that of the second.

Intra-layer $d$-SC is absent in the bilayer model, at least for the parameters that
we have considered and down to a temperature of $T=0.05$ (which appears to be sufficiently low to
allow extending this conclusion to the ground state). In accord with DMRG studies of Hubbard cylinders,
systems with $t'>0$ show stronger signatures of $d$-SC. However, none of them come close to the
required level for a superconducting instability. In contrast, we detect clear signs of inter-layer
spin-polarized superconductivity, which is expected in light of the attractive
ferromagnetic interaction that exists in the model.

Overall, our findings demonstrate that the bilayer model constitutes a valuable computational proxy to
the Hubbard model, and may serve as a controlled test bed to study further aspects of strongly correlated
electrons. To this end, one may consider augmenting the bilayer Hamiltonian with additional terms, which
nevertheless preserve the symmetry that keeps it free from the sign problem. This strategy may also be
pursued in order to suppress the inter-layer superconductivity since it can act as a masking agent that
obscures evidence for more interesting forms of superconductivity at low temperatures.


\begin{thebibliography}{999}
\bibitem{fitcomm} No background is added when fitting $S_s$. On purely phenomenological basis we use a background
of the form $\sqrt{|q_x/2\pi|}$ when fitting $S_c$. The position, width and height of the Lorentzians, as well as the
background amplitude, are obtained by a standard least squares fit.
\end{thebibliography}
\end{document}


\title{Supplemental Material for: "Quantum Monte Carlo study of a bilayer $U(2)\times U(2)$ symmetric Hubbard model"}

\author{Yosef Caplan$^{1}$ and Dror Orgad$^{1}$}
 \affiliation{$^1$Racah Institute of Physics, The Hebrew University, Jerusalem 91904, Israel}

\maketitle

\subsection{Sensitivity of the results to the value of $\Delta\tau$}

\begin{figure}[!!!h]
\begin{center}
\includegraphics[width = 0.492\textwidth]{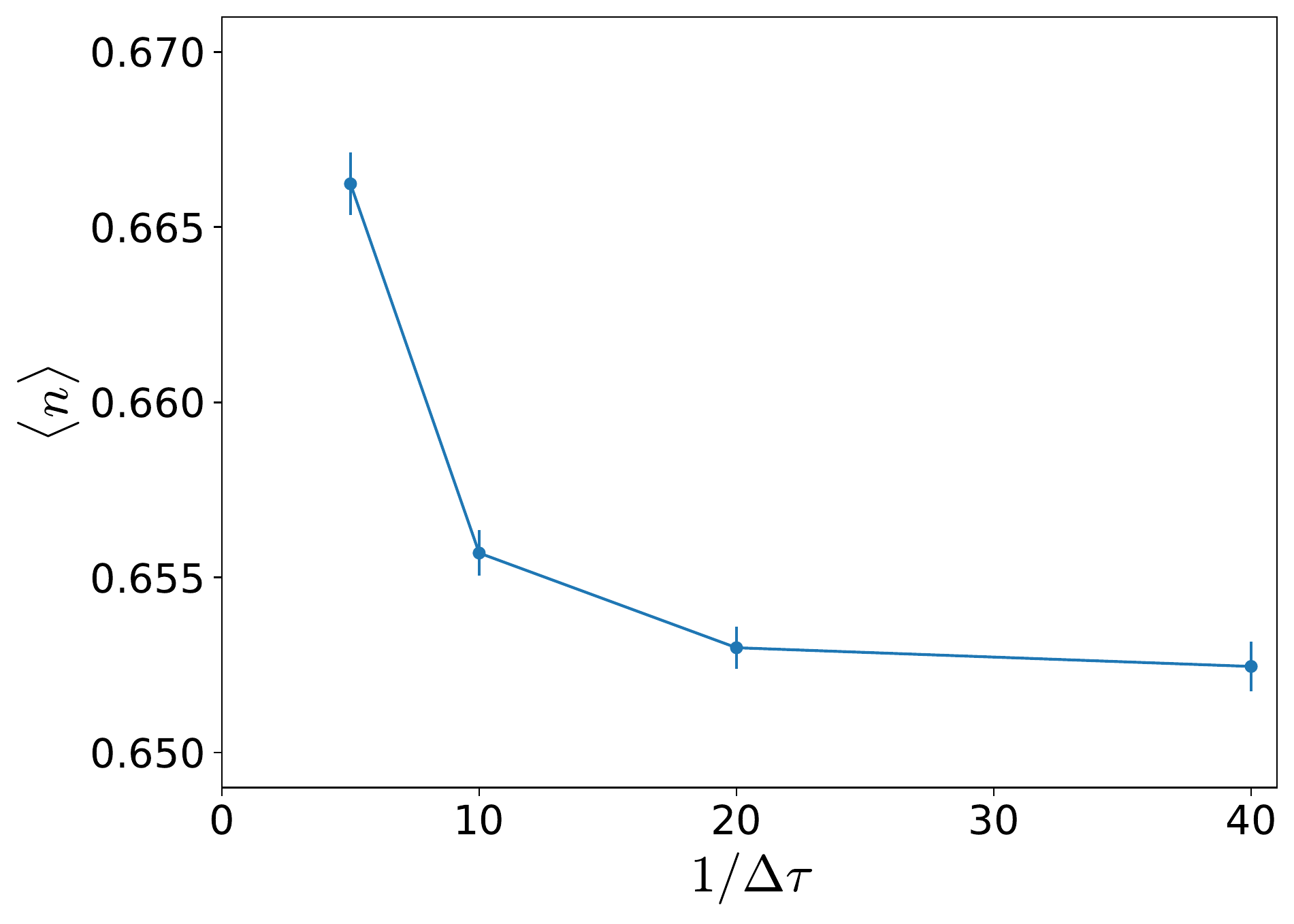}
\hspace{3mm}
\includegraphics[width = 0.48\textwidth]{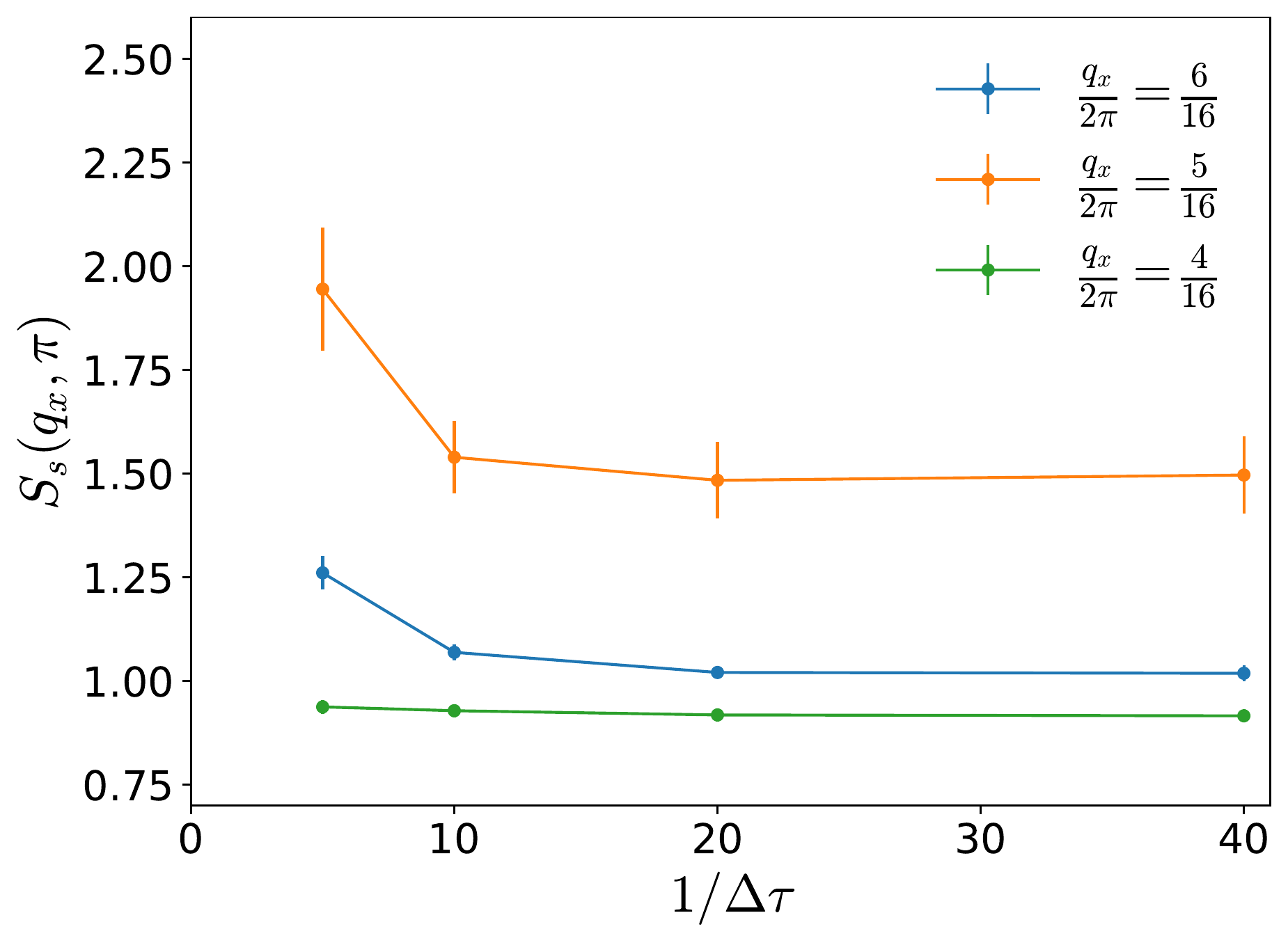}
\end{center}
\vspace{-0.5cm} \caption[]{\small Left: The dependence of the electronic density on $\Delta\tau$ in a system with $L=16$, $t'=0$ and $\beta=20$.
Right: The $\Delta\tau$ dependence of the $z$-spin structure factor at various values of $q_x$ in the same system with  $\mu=-1.7$. }
\label{fig-dtau}
\end{figure}

Fig. \ref{fig-dtau} depicts representative examples of the dependence of the electronic density and of the spin structure factor
on the value of the Trotter step size $\Delta\tau$. The figure demonstrates that decreasing $\Delta\tau$ beyond 0.1 (the value
used by us throughout the calculations) leads to a change of less than 1$\%$ in $\langle n\rangle$ and less than 5$\%$ in $S_s$.

\subsection{Dependence of the electronic density on the chemical potential}
\vspace{-0.75cm}
\begin{figure}[!!!h]
\begin{center}
\includegraphics[width = 0.52\textwidth]{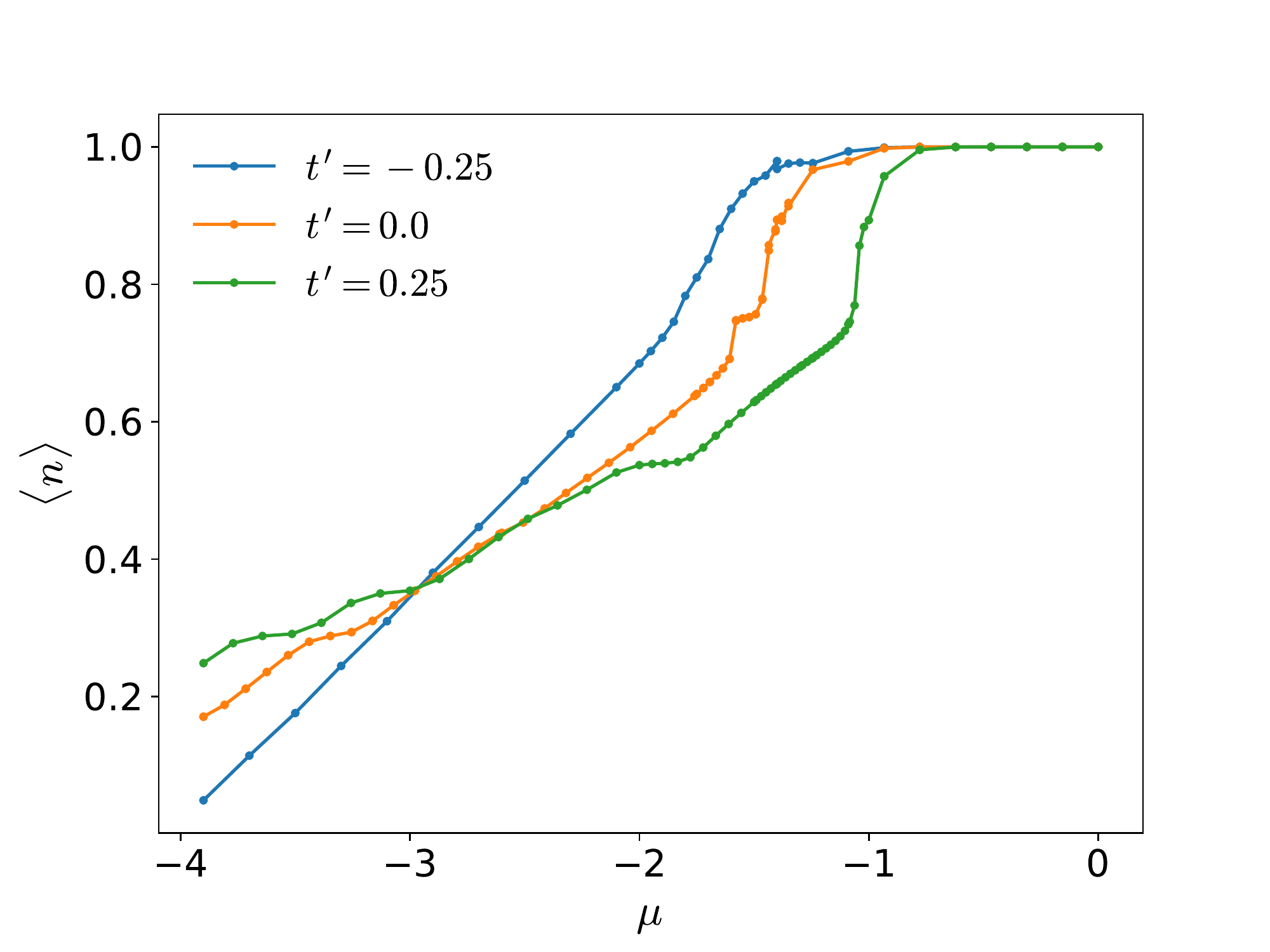}
\end{center}
\vspace{-0.5cm} \caption[]{\small The dependence of the electronic density on the chemical potential for $L=16$ systems at $\beta=20$.}
\label{fig-nmu}
\end{figure}

\newpage

\subsection{Examples of the spin susceptibility along the $z$ and transverse directions}
To make the consequences of breaking the spin-SU(2) symmetry apparent we include examples of the spin susceptibility
\begin{equation}
\chi_s^\mu(\bq)=2\int_0^\beta d\tau \sum_{l,i}e^{-i\bq\cdot \br_i}\langle S^\mu(l,\br_i,\tau)
S^\mu(l,{\bm 0},0)\rangle,
\end{equation}
along different directions.
Here $S^\mu(l,\br_i)=\frac{1}{2}\sum_{\alpha,\beta=\uparrow,\downarrow}c^\dagger_{l,\br_i,\alpha}\sigma^\mu_{\alpha\beta}c_{l,\br_i,\beta}$,
with ${\bm \sigma}$ the Pauli matrices.
\hspace{8mm}
\begin{figure}[!!!h]
\begin{center}
\includegraphics[width = 0.52\textwidth]{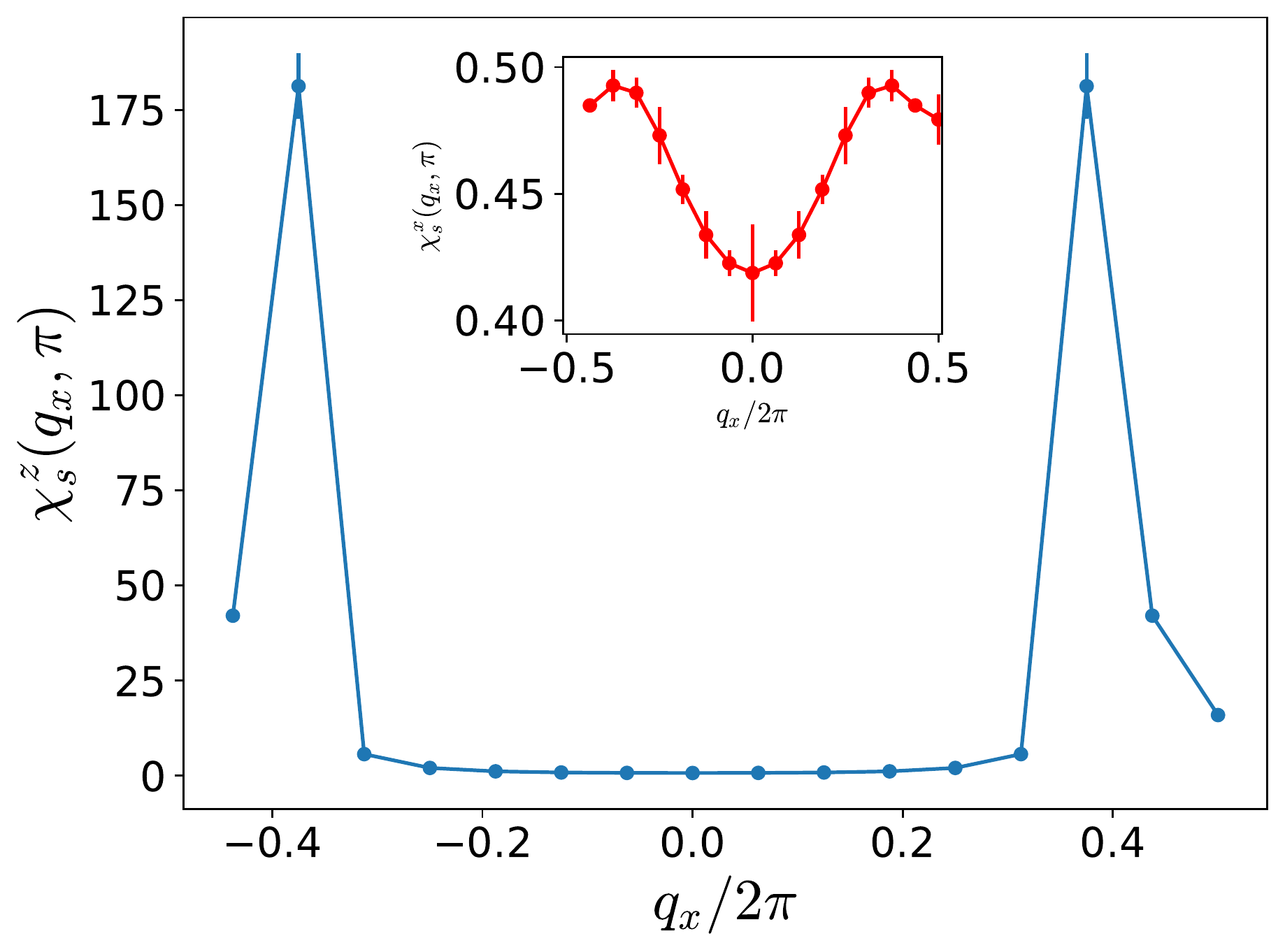}
\end{center}
\vspace{-0.5cm} \caption[]{\small The spin susceptibility in the $z$ and $x$ directions for a $L=16$, $t'=-0.25$ system at $\beta=20$  and $\delta=0.165$. }
\label{fig-chis}
\end{figure}

\newpage

\subsection{The inter-layer pairing susceptibility and the corresponding vertex}
Here we show results for the inter-layer pairing susceptibility
\begin{equation}
\chi_\perp=\frac{1}{2}\int_0^{\beta}d\tau \sum_{i,\sigma}\langle \Delta _{\perp\sigma}(\br_i,\tau)
\Delta_{\perp\sigma}^{\dagger}({\bm 0},0)\rangle,
\label{eq:chi_perp}
\end{equation}
and for $\Gamma_\perp\bar{\chi}_\perp$, where the inter-layer pairing vertex is defined via $\chi_\perp$ and the
uncorrelated inter-layer pair-field susceptibility $\bar{\chi}_\perp$ according to
\begin{equation}
\Gamma_\perp=\frac{1}{\chi_\perp}-\frac{1}{\bar{\chi}_\perp}.
\end{equation}

\vspace{1cm}
\begin{figure}[!!!h]
\begin{center}
\includegraphics[width = 0.46\textwidth]{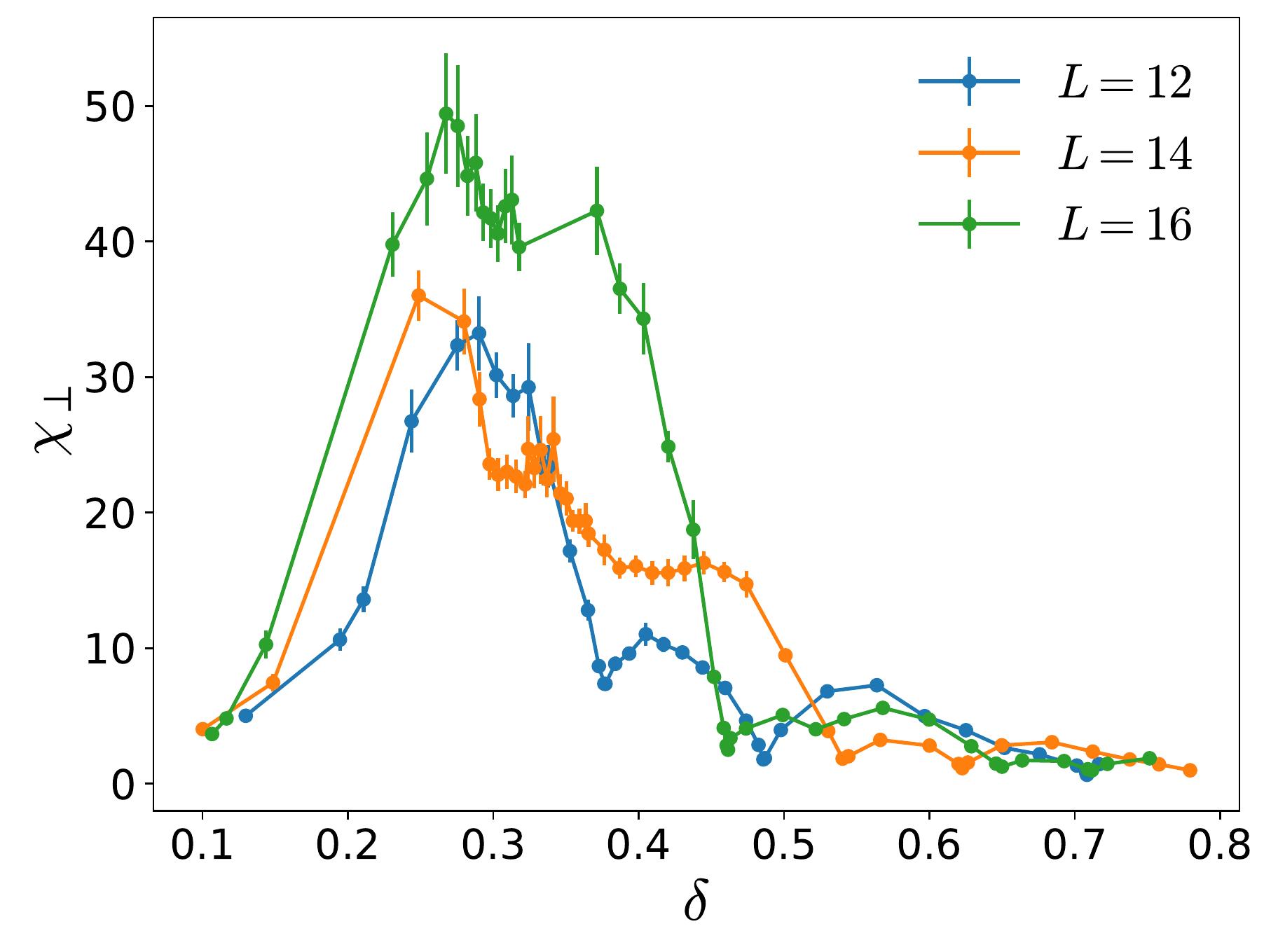}
\hspace{3mm}
\includegraphics[width = 0.475\textwidth]{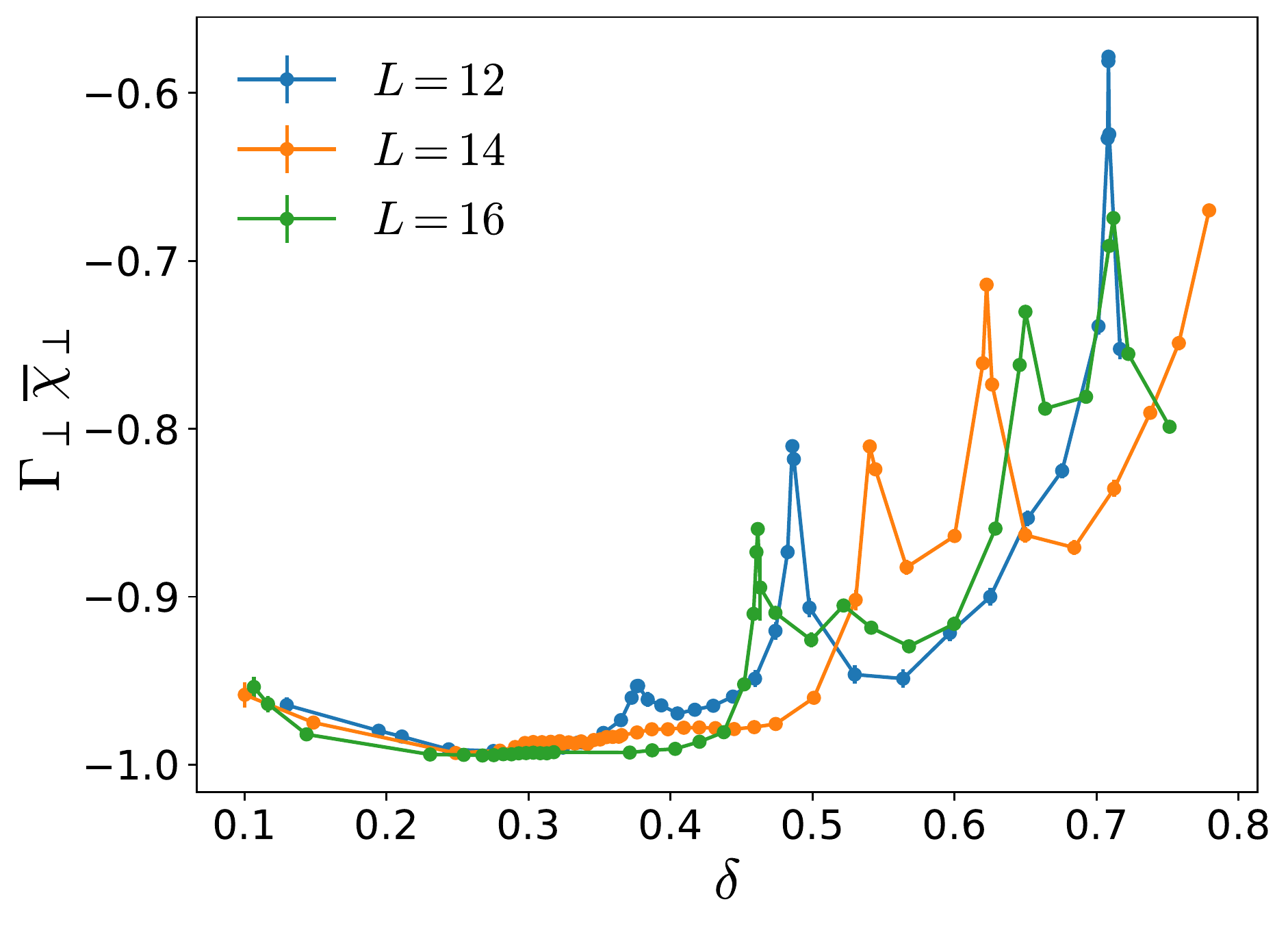}
\end{center}
\vspace{-0.5cm} \caption[]{\small Left: The inter-layer pairing susceptibility as a function of hole doping for square
periodic systems with $t'=0.25$ at $\beta=20$. Right: $\Gamma_\perp\bar{\chi}_\perp$ for the same systems.}
\label{fig-chiperp}
\end{figure}